\documentclass[twocolumn,amsmath,amssymb]{revtex4} 

\usepackage{graphicx}
\usepackage{dcolumn}
\usepackage{bm}
\usepackage{braket}

\begin{document}

\title{Experimental Examination of the Effect of Short Ray Trajectories in Two-port Wave-Chaotic Scattering Systems}

\author{Jen-Hao Yeh$^{1}$}
\author{James A. Hart$^{1,2}$}
\author{Elliott Bradshaw$^{2}$}
\author{Thomas M. Antonsen$^{1,2}$}
\author{Edward Ott$^{1,2}$}
\author{Steven M. Anlage$^{1,2}$}

\affiliation{$^{1}$Department of Electrical and Computer
Engineering, University of Maryland, College Park, MD 20742-3285}

\affiliation{$^{2}$Physics Department, University of Maryland,
College Park, MD  20742-4111}

\date{\today}

\begin{abstract}
\section*{Abstract}
Predicting the statistics of realistic wave-chaotic scattering
systems requires, in addition to random matrix theory, introduction
of system-specific information. This paper investigates
experimentally one aspect of system-specific behavior, namely the
effects of short ray trajectories in wave-chaotic systems open to
outside scattering channels. In particular, we consider ray
trajectories of limited length that enter a scattering region
through a channel (port) and subsequently exit through a channel
(port). We show that a suitably averaged value of the impedance can
be computed from these trajectories and that this can improve the ability
to describe the statistical properties of the scattering systems. We
illustrate and test these points through experiments on a realistic
two-port microwave scattering billiard.
\end{abstract}
\maketitle

\section*{PACS Numbers}
05.45.Mt Quantum chaos; semiclassical methods

03.65.Nk Scattering theory

03.65.Sq Semiclassical theories and
applications

05.60.Gg Quantum transport

\section{Introduction}
Random matrix theory (RMT) has achieved substantial success at
predicting short wavelength statistical properties of spectra,
eigenfunctions, scattering matrices, impedance matrices, and
conductance of wave-chaotic systems \cite{ Verb,Mehta,
Beenakker_review_RMT, Stock, Fyodorov}. By wave-chaotic systems, we
mean that the behavior of the wave system in the small wavelength
limit is described by ray orbit trajectories that are chaotic
\cite{S1}. In practice, the experimental applicability of RMT
requires consideration of nonuniversal effects. For example, in the
particular case of scattering, the scattering properties of an open
system depend on the coupling between the field within the
scattering region and the asymptotic incoming and outgoing waves
connecting the exterior to the scatterer. For this problem,
researchers have developed methods to incorporate nonuniversal
coupling and port-specific effects into the analysis of short
wavelength scattering data for systems whose closed classical
counterparts are ray-chaotic \cite{Poisson_Kernel_Original,
Brouwer_Lorentzian, Poisson_including_internal, Kuhl, S1,
Henry_paper_one_port, Henry_paper_many_port, Richter, Haake}.

Previous comparisons of experimental data to RMT have often employed
ensembles of realizations of the system to compile statistics. To
create such ensembles, researchers have typically varied the
geometrical configuration of the scattering region and/or taken
measurements at several different wavelengths
\cite{Henry_paper_one_port, Henry_paper_many_port, S1,Schafer}.
These variations aim to create a set of systems in which none of the
nonuniversal system details are reproduced from one realization to
another, except for the effects of the port details. Thus, by
suitably accounting for the port details, it was hoped that only
universal RMT properties remained in the ensemble data. However,
there can be problems in practice. For example, in the case of
geometrical configuration variation, researchers typically move
perturbing objects inside a ray-chaotic enclosure with fixed shape
and size \cite{ S1,Henry_paper_one_port, Henry_paper_many_port}, or
move one wall of that enclosure \cite{Schafer}, to create an
ensemble of systems with varying details. The problem is that
certain walls or other scattering objects of the enclosure remain
fixed throughout the ensemble. Therefore, there may exist relevant
ray trajectories that remain unchanged in many or all realizations
of the ensemble. We term such ray trajectories that leave a port and
soon return to it (or another port) before ergodically sampling the
enclosure ``short ray trajectories''.

A similar problem arises for wavelength variation in which a band of
wavelengths is used. Within a wavelength range, if a ray trajectory
length is too short, then the variation of the phase accumulated by
a wave following that trajectory may not be large enough to be
considered random. In such a case the effects of specific (hence,
nonuniversal) short ray trajectories will survive the ensemble
averaging processes described above. These problems will make
systematic, nonuniversal contributions to the ensemble data, and
thus consideration of short ray trajectories arises naturally in the
semiclassical approach to quantum scattering theory
\cite{wirtz,Richter, Haake, Nazmitdinov,Stampfer,ishio_and_burgdoerfer,Stein,ishio_and_keating,prange_alone,gutzwiller_book,
Stock}. Such short-ray-trajectory contributions have been noted
before in microwave billiards \cite{Henry_paper_one_port,
Henry_paper_many_port, ZSO} or for quantum transport in chaotic cavities \cite{Richter, Haake}
and have either constrained or frustrated previous tests of RMT predictions.

By a ``short ray trajectory'' we mean one whose length is not much
longer than several times the characteristic size of the scattering
region, and which enters the scattering region from a port, bounces
(perhaps several times) within the scattering region, and then
returns to a port. A ``port'' is the region in which there is a
connection from the scatterer to the outside world. For illustrative
purposes, in what follows we consider systems with either one or two
ports; as discussed elsewhere, generalization of our results to more
ports is straightforward \cite{ZSO, S3, Experimental_tests_our_work,
Hart}. Note that the short ray trajectories we refer to are
different from periodic orbits \cite{Sieber, Stein, Wintgen}, which are
closed classical trajectories; short ray trajectories, as defined
here, are only important for open systems. Although our explicit
considerations in this paper are for billiard systems (i.e.,
scattering regions that are homogeneous with perfectly reflecting
walls), presumably these effects may also be present with continuous
potentials and are not limited to billiards.

In this paper we go beyond the previous treatments to explicitly
include additional nonuniversal effects due to short ray
trajectories. Previous work has examined short ray trajectories in
cases where the system and the ports can be treated in the
semiclassical approximation \cite{ishio_and_burgdoerfer,
prange_alone} or considered the effect on eigenfunction correlations
due to short ray trajectories associated with nearby walls
\cite{urbina_richter_semiclassical_construction,urbina_richter_beyond_universality}.
The effects of short ray trajectories on wave scattering properties
of chaotic systems have been explicitly calculated before in the
case of quantum graphs \cite{Kottos} and for two dimensional
billiards \cite{Hart}. Moreover, microwave billiard experimental
work has extracted a measure of the microwave power that is emitted
at a certain point in the billiard and returns to the same point
after following all possible classical trajectories of a given
length \cite{Stein}. The Poisson Kernel approach can also be
generalized to include short ray trajectories
\cite{Bulgakov_Gopar_Mello_Rotter} through measurement of a
(statistical) optical S-matrix. None of this prior work developed a
general first-principles deterministic approach to experimentally
analyzing the effect of short ray trajectories, as we do here. This
paper expands on a brief report of our preliminary results
\cite{Jenhao}. Our previous work \cite{Jenhao} demonstrated the
effect of short trajectories in a one-port wave-chaotic cavity, and
we generalize the short-trajectory correction to a two-port system
in this paper. Therefore, this generalized correction can be applied
to multiple port cases by similar procedures. Other new ingredients
include detailed examination of the effect of individual short ray
trajectories, and extension to the statistics of other wave
scattering properties.

The outline of this paper is as follows. In Sec. \ref{sec:theory},
we summarize a theoretical approach for removing short ray
trajectories from single-realization data and ensemble-averaged data
of any wave properties of a wave scattering system. A detailed
derivation of the results in Sec. \ref{sec:theory} is given
elsewhere \cite{Hart}. In Sec. \ref{sec:experiment}, we describe
experiments testing the theoretical approach, and we compare the
experimental results and the theory in Sec. \ref{sec:results}. These
comparisons show that compensation for the effects of short ray
trajectories improves the agreement between RMT-based predictions
and measured statistical properties of ensemble-averaged data and
single-realization data.

While much of our discussion above has emphasized the goal of
uncovering RMT statistics, we also wish to emphasize that many of
our results are also of interest independent of that goal. In
particular, the key step in our approach for uncovering RMT
statistics is that of obtaining a short-trajectory-based prediction
for the average impedance over frequency and/or configurations
(Secs. \ref{subsec:empty_cavity} and \ref{subsec:ensemble}), and we
note that this quantity is both experimentally accessible and of
interest in its own right. As another example, in Sec.
\ref{subsec:wall_cases} we consider the modification of the
free-space radiation impedance arising in configurations when there
are only a few possible short ray trajectories, and this situation
applies directly to many cases where, although reflecting objects
are present, there is also substantial coupling to the outside.

\section{Review of Theory}\label{sec:theory}
In this paper we consider for concreteness an effectively two
dimensional microwave cavity (described in Sec.
\ref{sec:experiment}) as the scattering system, and this cavity is
connected to the outside world via one or two single-mode
transmission lines terminated by antennas inside the scattering
region, acting as ports. The results and techniques should carry
over to other physically different wave systems (e.g., quantum
waves, acoustic waves, etc.) of higher dimension and an arbitrary
number of ports \cite{Hart}. Ray trajectories within the cavity are
chaotic due to the shape of the cavity.

Previously most researchers focused on the scattering matrix $S$
\cite{Dyson_original, Poisson_Kernel_Original,
Doron_Smilansky_Poisson, Poisson_including_internal,
Poisson_including_internal_2, Barthelemy1, Barthelemy2, A_Richter,
Dietz}, which specifies the linear relationship between reflected
and incident wave amplitudes in the channels. We focus on an
equivalent quantity, the impedance $Z=Z_{0}(1+S)/(1-S)$, because
nonuniversal contributions manifest themselves in $Z$ as simple
additive corrections \cite{impedance}. Here the diagonal matrix
$Z_0$ gives the characteristic impedance of the scattering channels,
and is measured at chosen reference planes on the transmission lines
as the ratio of the (complex) transmission line voltage $\hat{V}$ to
current $\hat{I}$ (time variation $exp(-i\omega t)$ is assumed).
Impedance is a meaningful concept for all scattering wave systems.
In linear electromagnetic systems, it is defined via the phasor
generalization of Ohm's law as $\hat{V}=Z\hat{I}$, and in the case
of $N$ channels connected to the scatterer, $Z$ is an $N\times N$
matrix. A quantum-mechanical quantity corresponding to the impedance
is the reaction matrix, which is often denoted in the literature as
$K=-iZ$ \cite{Verb, Fyodorov, S3, Alhassid_review,
Lossy_approximate_impedance_Savin_Fyodorov,
Lossy_impedance_Savin_Sommers_Fyodorov, henrythesis}. In what
follows our discussion will use language appropriate to the
electromagnetic context and scattering from a microwave cavity
excited by small antennas fed by transmission lines (the setting for
our experiments).

Our goal is to describe the universal RMT aspects of measurement of
this impedance matrix (alternatively of the scattering matrix).
However, raw measurement of the impedance have nonuniversal
properties because the waves are coupled to the cavity through the
specific geometry of the junction between the transmission lines and
the cavity. In prior work, the nonuniversal coupling effects were
parameterized in terms of corrections to the impedance $Z$. The
``perfectly coupled'' normalized impedance $i\xi_{0}$ removes
nonuniversal features due to the ports
\cite{Henry_paper_one_port,Henry_paper_many_port},
\begin{equation}\label{eq:normalized_impedance}
    i\xi_{0}=R_{R}^{-1/2}(Z-i X_{R})R_{R}^{-1/2},
\end{equation}
where $Z_{R}=R_{R}+iX_{R}$ is the radiation impedance, embodying the
nonuniversal port properties. $R_{R}$ is the real part of $Z_{R}$,
and $X_{R}$ is the imaginary part. By ``perfectly coupled'' we mean
that waves impinging on the cavity are fully transmitted to the
cavity with no prompt reflection. Note that these quantities are
$N\times N$ matrices for a system with $N$ ports. $Z_{R}$ is the
impedance measured at the previously mentioned reference planes on
the transmission lines when the cavity walls are removed, so that
the outgoing waves launched at the ports never return to these
ports. Eq. (\ref{eq:normalized_impedance}) removes the effect of the
ports in the impedance, so the ``perfectly coupled'' normalized
impedance
$i\xi_{0}$ of the cavity is all that remains. 
We note that $Z_{R}$ is experimentally accessible through a
deterministic (i.e., non-statistical) measurement described in Sec.
\ref{sec:experiment}.

In Hart et al. \cite{Hart}, it has been proposed that nonuniversal
effects due to the presence of short ray trajectories can be removed
by appropriate modification of the radiation impedance
$Z_{R}\rightarrow Z_{avg}=Z_{R}+$[short trajectory terms], where
$Z_{avg}\equiv Z_{0}(1+\bar{S})/(1-\bar{S})$ stands for the
``average'' impedance. $\bar{S}$ is the average scattering matrix,
and the Poisson kernel characterizes the statistical distribution of
the scattering matrix $S$ in terms of $\bar{S}$
\cite{Poisson_Kernel_Original, Doron_Smilansky_Poisson, Kuhl}. More
specifically, the averaged impedance is written as
\begin{equation}\label{eq:average_impedance}
    Z_{avg}=Z_{R}+R_{R}^{1/2}z R_{R}^{1/2},
\end{equation}
where $z$ is the short-trajectory correction. With this
modification, $i\xi_{0}$ in Eq. (\ref{eq:normalized_impedance}) is
extended to a perfectly-coupled and short-trajectory-corrected
normalized impedance
\begin{equation}\label{eq:SOC normalized impedance}
    i\xi=R_{avg}^{-1/2}\left(Z-iX_{avg}\right)R_{avg}^{-1/2},
\end{equation}
where $Z_{avg}=R_{avg}+iX_{avg}$. We define $z\equiv\rho+i\chi$, so
$R_{avg}$ and $X_{avg}$ can be written as
\begin{equation}\label{eq:real average_impedance}
    R_{avg}=R_{R}+R_{R}^{1/2}\rho R_{R}^{1/2},
\end{equation}
\begin{equation}\label{eq:imaginary average_impedance}
    X_{avg}=X_{R}+R_{R}^{1/2}\chi R_{R}^{1/2}.
\end{equation}

In the lossless case $\rho$ and $\chi$ are the real and imaginary
parts of $z$; $R_{avg}$ and $X_{avg}$ are the real and imaginary
parts of $Z_{avg}$. However, with uniform loss (e.g., due to an
imaginary part of a homogeneous dielectric constant in a microwave
cavity), $R_{avg}$ and $X_{avg}$ are the analytic continuations of
the real and imaginary parts of the lossless $Z_{avg}$ as
$k\rightarrow k+ik/(2Q)$, where $Q\gg 1$ is the quality factor of
the closed system, and $k$ denotes the wavenumber of a plane wave.
These analytic continuations are no longer purely real (i.e.,
$\rho$, $\chi$, $R_{avg}$ and $X_{avg}$ become complex). The
normalized impedance $i\xi$ of the lossy system has a universal
distribution which is dependent only on the ratio $k/(2Q \Delta k)$,
where $\Delta k$ is the mean spacing between modes \cite{Hart,
Lossy_approximate_impedance_Savin_Fyodorov,
Lossy_impedance_Savin_Sommers_Fyodorov, Fyodorov,
Henry_paper_one_port, Henry_paper_many_port}.

For the system with $N$ ports, the elements of the matrices $z$ are
\cite{Hart}
\begin{equation*}
    z_{n,m}=\sum_{b(n,m)}\{-p_{b(n,m)}\sqrt{D_{b(n,m)}}\exp[-(ik+\kappa)L_{b(n,m)}
\end{equation*}
\begin{equation}\label{eq:SOC}
    -ikL_{port(n,m)}-i\beta_{b(n,m)}\pi]\}.
\end{equation}
With $\rho_{n,m}+i\chi_{n,m} \equiv z_{n,m}$, the analytic
continuations of $\rho_{n,m}$ and $\chi_{n,m}$ are
\begin{equation*}
    \rho_{n,m}=\sum_{b(n,m)}\{-p_{b(n,m)}\sqrt{D_{b(n,m)}}\cos[-(k-i\kappa)L_{b(n,m)}
\end{equation*}
\begin{equation}\label{eq:real part SOC}
    -kL_{port(n,m)}-\beta_{b(n,m)}\pi]\},
\end{equation}
\begin{equation*}
    \chi_{n,m}=\sum_{b(n,m)}\{-p_{b(n,m)}\sqrt{D_{b(n,m)}}\sin[-(k-i\kappa)L_{b(n,m)}
\end{equation*}
\begin{equation}\label{eq:imaginary part SOC}
    -kL_{port(n,m)}-\beta_{b(n,m)}\pi]\},
\end{equation}
where $b(n,m)$ is an index over all classical trajectories which
leave the \textit{n}th port, bounce $\beta_{b(n,m)}$ times, and
return to the \textit{m}th port, $L_{b(n,m)}$ is the length of the
trajectory $b(n,m)$, $\kappa=k/(2Q)$ is the effective attenuation
parameter taking account of loss, and $L_{port(n,m)}$ is the
port-dependent constant length between the \textit{n}th port and the
\textit{m}th port. $D_{b(n,m)}$ is a geometrical factor of the
trajectory, and it takes into account the spreading of the ray tube
along its path. This geometrical factor is a function of the length
of each segment of the trajectory, the angle of incidence of each
bounce, and the radius of curvature of each wall encountered in that
trajectory; it has been assumed that the port radiates isotropically
from a location far from the two-dimensional cavity boundaries.
$p_{b(n,m)}$ is the survival probability of the trajectory in the
ensemble and will be discussed subsequently. In the lossless case
($\kappa=0$), note that $\rho$ and $\chi$ are both real, but they
are both complex in the presence of loss. In arriving at Eqs.
(\ref{eq:real part SOC}) and (\ref{eq:imaginary part SOC}) it is
assumed that the lateral walls present perfect-metal boundary
conditions, and that foci and caustics are absent. These assumptions
are well satisfied for the cavity shape used in our experiments.

Note that the sums in Eqs. (\ref{eq:real part SOC}) and
(\ref{eq:imaginary part SOC}) involve terms of the form sine and
cosine of $[(k-i\kappa)L_{b(n,m)} + \ldots]$ which for large
$L_{b(n,m)}$ increase exponentially like $\exp(\kappa L_{b(n,m)} )$.
Although in the experiments we have observed that the parameters
$p_{b(n,m)}$ and $D_{b(n,m)}$ decrease exponentially when
$L_{b(n,m)}$ increases, the sums do not necessarily converge if the
loss is too high. Accordingly, we will use a finite cutoff of the
sum and regard the cutoff result as being asymptotic. In contrast,
if instead of extracting RMT statistics, we regard our goal as
approximating $Z_{avg}=R_{avg}+iX_{avg}$, then Eq.
(\ref{eq:average_impedance}) shows that we only need to consider
$z_{n,m}$. Thus the sum involved in the calculation of $z$ (Eq.
(\ref{eq:SOC})) is now over terms that \textit{decrease}
exponentially with increasing path length as $\exp(-\kappa
L_{b(n,m)})$. This sum is much more likely to converge than the sums
in Eqs. (\ref{eq:real part SOC}) and (\ref{eq:imaginary part SOC}).

In practice, when considering either of the sums in Eqs.
(\ref{eq:SOC}), (\ref{eq:real part SOC}) or (\ref{eq:imaginary part
SOC}), we employ a cutoff by replacing the sums by
$\displaystyle\sum_{b(n,m)}^{L_{M}}$ which signifies that the sum is
now over all trajectories $b(n,m)$ with lengths up to some maximum
length $L_{M}$, $L_{b(n,m)}\leq L_{M}$, and this is the reason of
the name ``short-trajectory correction''. We use $\rho^{(L_{M})}$,
$\chi^{(L_{M})}$, $z^{(L_{M})}$, $i\xi^{(L_{M})}$,
$Z^{(L_{M})}_{avg}$, $R^{(L_{M})}_{avg}$, and $X^{(L_{M})}_{avg}$ to
indicate the finite length versions of those quantities. Combined
with the measured radiation impedance, these corrections can be
analytically determined by using the ray-optics of short ray
trajectories between the ports and the fixed walls of the cavity.
After understanding the nonuniversal effects of radiation impedance
and short trajectories, we compare the predictions of RMT with the
normalized impedance $i\xi$ which contains the remaining features of
longer trajectories and the deviations between a single realization
and the ensemble average.

In Sec. \ref{sec:experiment}, in addition to configuration averaging
[taken into account by the quantity $p_{b(n,m)}$ in Eqs.
(\ref{eq:SOC}), (\ref{eq:real part SOC}) or (\ref{eq:imaginary part
SOC}) (see Sec. \ref{sec:experiment})], we will also employ
frequency smoothing. In particular, if $f(\omega)$ denotes a
frequency dependent quantity, then we take its frequency smoothed
counterpart to be the convolution of $f(\omega)$ with a gaussian,
\begin{equation}\label{eq:Gaussian_smoothing}
    \overline{f(\omega)} = \int f(\omega')g(\omega-\omega')d\omega',
\end{equation}
where
\begin{equation}\label{eq:Gaussian_function}
    g(\omega)=\frac{1}{\sqrt{2\pi}\Delta\omega}\exp\left(\frac{-\omega^{2}}{2(\Delta\omega)^{2}}\right).
\end{equation}
Applying the operation (\ref{eq:Gaussian_smoothing}) to our
short-trajectory correction formulas, Eqs. (\ref{eq:SOC}),
(\ref{eq:real part SOC}) or (\ref{eq:imaginary part SOC}), with
$k=\omega/c$, we see that the summations acquire an additional
multiplicative factor,
\begin{equation}\label{eq:Gaussian_factor}
   \exp[-\frac{1}{2}(L_{b(n,m)}+L_{port(n,m)})^{2}(\Delta\omega/c)^{2}].
\end{equation}
Thus, as $L_{b(n,m)}$ increases (i.e., longer trajectories are
included), the factor (\ref{eq:Gaussian_factor}) eventually becomes
small, thus providing a natural cutoff to the summations in Eqs.
(\ref{eq:real part SOC}), (\ref{eq:imaginary part SOC}) and
(\ref{eq:SOC}).

\section{Experiment}\label{sec:experiment}
We have carried out experimental tests of the short-trajectory
effects using a quasi-two-dimensional microwave cavity with two
different port configurations, a one-port system and a two-port
system. For the one-port case $Z$ and $S$ are scalars, while they
are $2\times 2$ matrices in the two-port case. Microwaves are
injected through each antenna attached to a coaxial transmission
line of characteristic impedance $Z_{0}$, and the antennas are
inserted into the cavity through small holes (diameters about 1 mm)
in the lid, similar to previous setups \cite{S1,S2,S3,
Experimental_tests_our_work}. The waves introduced are
quasi-two-dimensional (cavity height 0.8 cm) for frequencies (from 6
to 18 GHz) below the cutoff frequency for higher order modes ($\sim
19$ GHz), including about 1070 eigenmodes of the closed cavity. The
quasi-two-dimensional eigenmodes of the closed system are described
by the Helmholtz equation for the single non-zero component of
electric field ($E_{z}$), and these solutions can be mapped onto
solutions of the Schr\"{o}dinger equation for an infinite square
well potential of the same shape \cite{Stock, Ali_Gokirmak}.

The shape of the cavity walls is a symmetry-reduced ``bow-tie
billiard'' made up of two straight walls and two circular dispersing
walls \cite{Ali_Gokirmak}, as shown in the insets of Fig.
\ref{fig:walls_case}. The lengths of wall A and wall B are 21.6 cm
and 43.2 cm, respectively, while the wavelength of the microwave
signals ranges from 1.7 cm to 5.0 cm, putting this billiard system
just into the semiclassical regime. Despite the fact that we are not
deep inside the semiclassical regime, we obtain results in good
agreement with theory (Sec. \ref{sec:results}). In the one-port
experiments the location of the port is 18.0 cm from wall A and 15.5
cm from wall B. In the two-port experiments we add the other port at
the location 35.6 cm from wall A and 15.6 cm from wall B. The cavity
shape yields chaotic ray trajectories and has been previously used
to examine eigenvalue \cite{So} and eigenfunction \cite{Chung}
statistics of the closed system in the crossover from GOE to GUE
statistics as time-reversal invariance is broken. In addition, this
cavity was also used to study scattering ($S$-), \cite{S2, S3}
impedance ($Z$-) \cite{S1} and conductance ($G$-)
\cite{Experimental_tests_our_work} statistics.

We use an Agilent PNA network analyzer to measure the frequency
dependence of the complex scattering matrix $S$, and compute the
corresponding impedance $Z=Z_{0}(1+S)/(1-S)$. The walls are fixed
relative to the ports in all experiments. We experimentally
determine the radiation impedance $Z_{R}$ by placing microwave
absorbers along all the side-walls of the cavity and measuring the
resulting impedance. The absorbers eliminate reflections from the
walls, so this method removes the effect of ray trajectories,
leaving only the effects of the port details. To verify the theory,
in preliminary experiments microwave absorbers are placed only along
specific walls, and we measure and examine specific contributions of
individual walls, or groups of walls, to the impedance (Sec.
\ref{subsec:wall_cases} and \ref{subsec:errors}). In another set of
experiments, we remove all the absorbers and make measurement on the
empty cavity (Sec. \ref{subsec:empty_cavity}). Finally, we also make
measurement with two perturbers added to the interior of the cavity
(Sec. \ref{subsec:ensemble}). Each perturber is a cylindrical piece
of metal with a height similar to that of the cavity, and their
function is to block ray trajectories. The two perturbers are
systematically moved to create 100 different realizations used to
form an ensemble of systems in which the empty-cavity short
trajectories are partially destroyed. In the one-port experiments
the cross sections of the two cylindrical perturbers are
irregularly-starlike shapes with the maximum diameters 7.9 and 9.5
cm. In the two-port experiments the cross sections are identical
circular shapes with diameter 5.1 cm.

\section{Results}\label{sec:results}
\subsection{Individual short ray trajectories}\label{subsec:wall_cases}
\begin{figure}
\includegraphics[height=1.8in,width=2.2in]{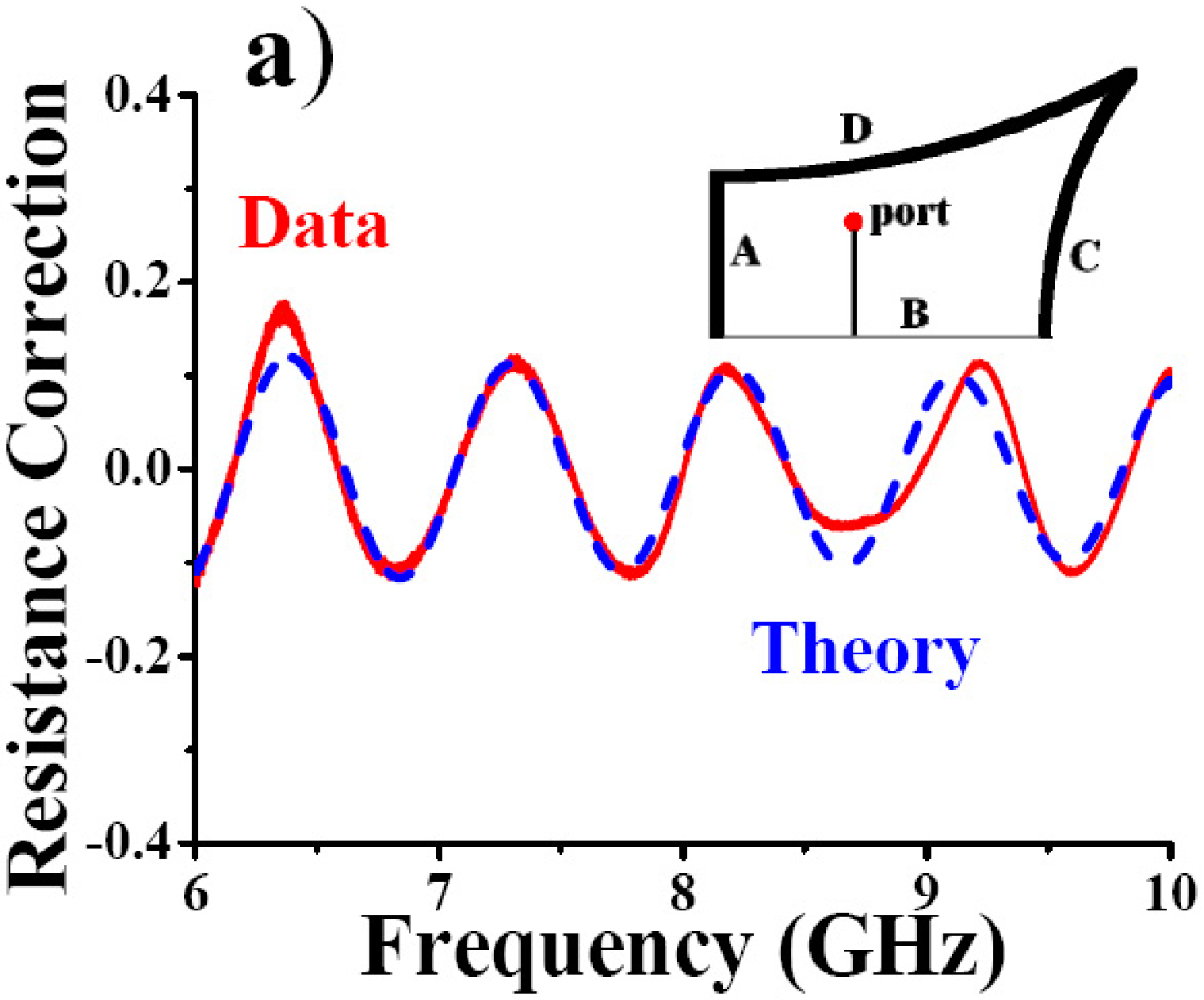}
\includegraphics[height=1.8in,width=2.2in]{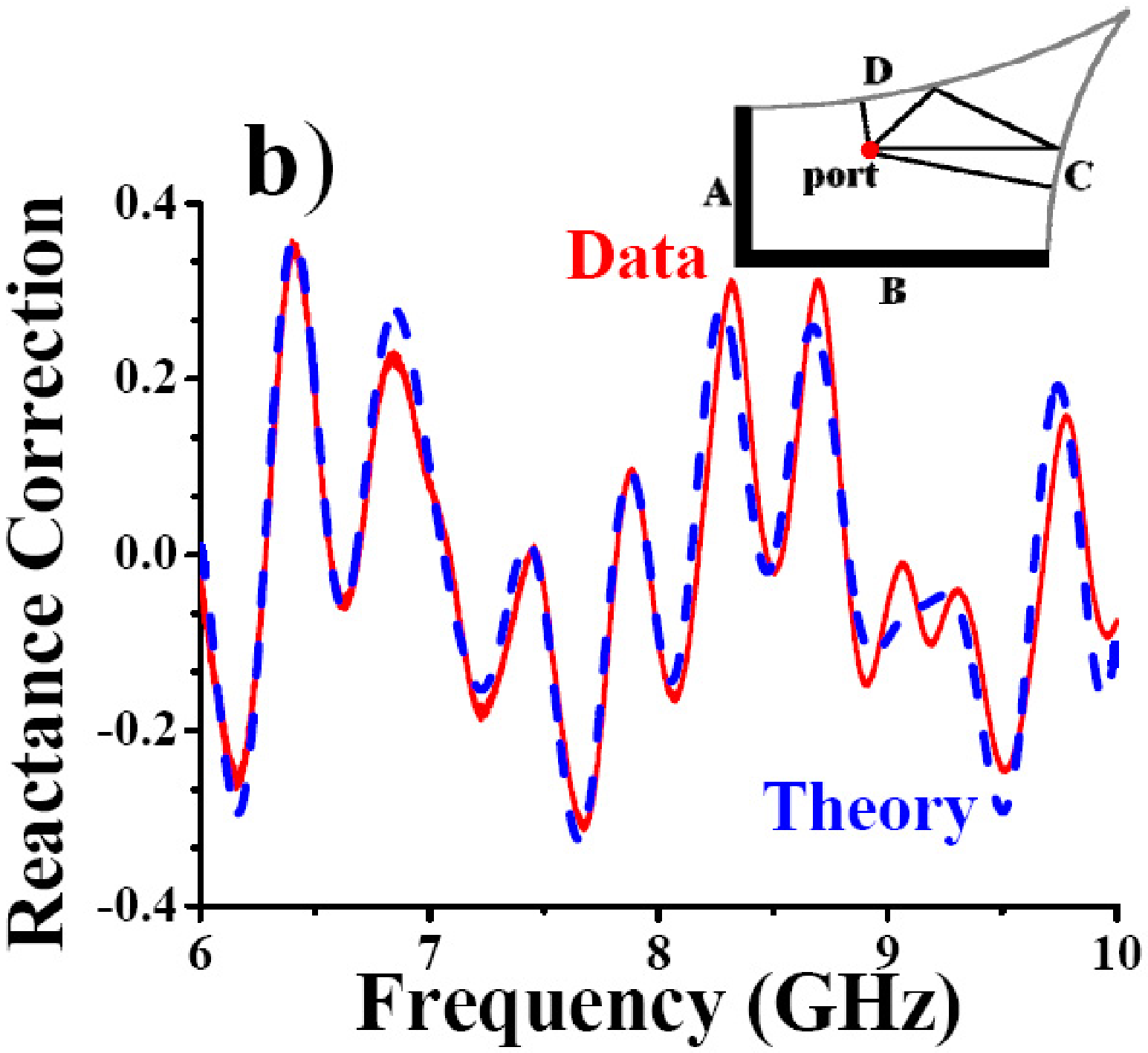}
\includegraphics[height=1.8in,width=2.2in]{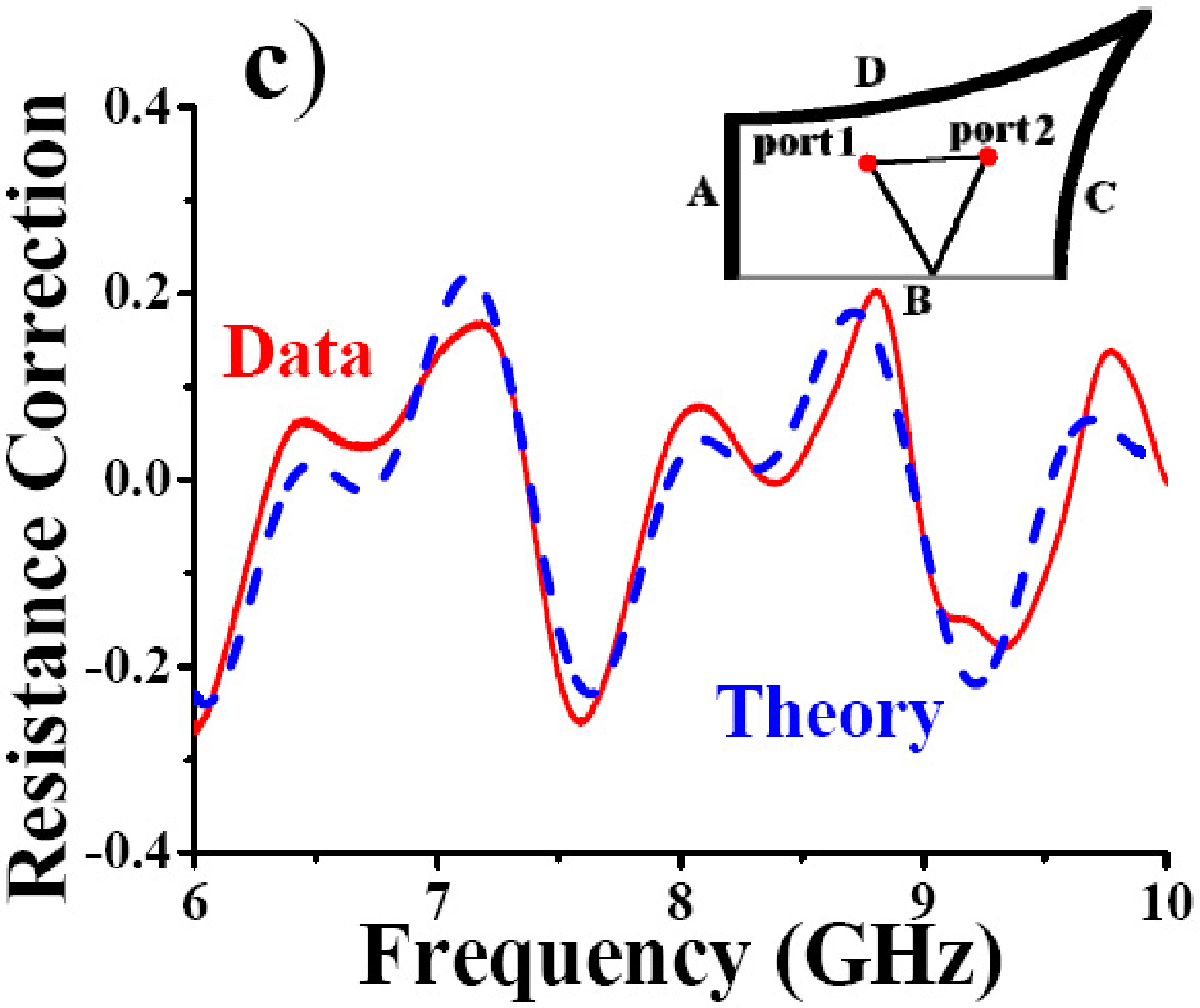}
\includegraphics[height=1.8in,width=2.2in]{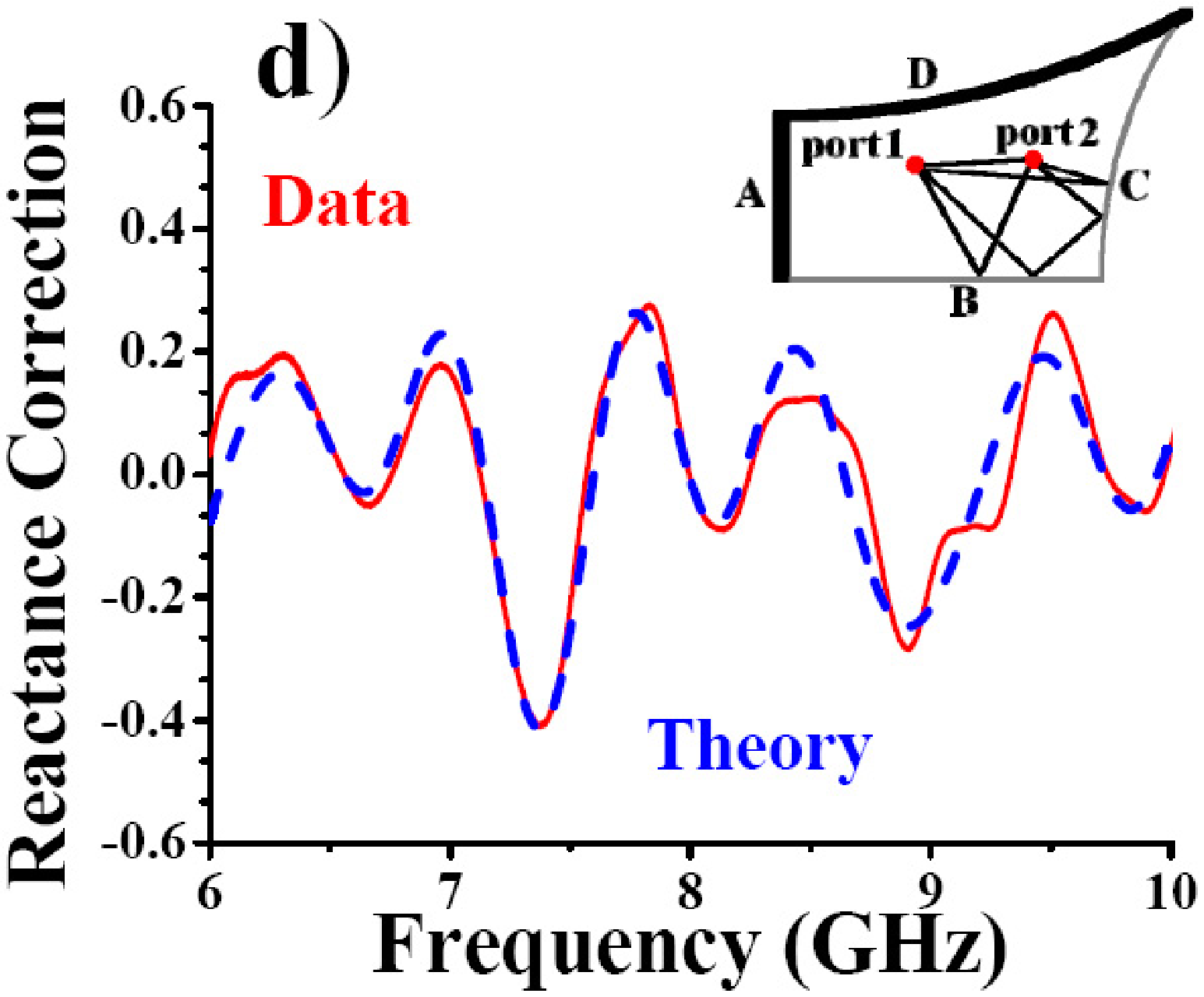}
\caption{(Color online) Plot of corrections to the impedance due to
(a) one-port with one wall (wall B), (b) one-port with two walls
(walls C and D), (c) two-port with one wall (wall B), and (d)
two-port with two walls (walls B and C) short trajectories, versus
frequency from 6 to 10 GHz inside the $1/4$-bow-tie cavity. Shown
are the theoretical predictions ($\textrm{Re}[z_{cor}^{(L_{M})}]$
and $\textrm{Im}[z_{cor}^{(L_{M})}]$) in blue (dashed) curves and
experimental data ($\textrm{Re}[z_{cor}]$ and
$\textrm{Im}[z_{cor}]$) in red (solid) curves. \\
The insets show the $1/4$-bow-tie billiard with thicker lines
representing walls covered by microwave absorbers, the ports are
shown as circular dots, and a few short trajectories are represented
with lines as illustrations.} \label{fig:walls_case}
\end{figure}

The first experiment tests whether the theory of short-trajectory
corrections can predict the effect of individual ray trajectories,
as well as the aggregate effect of a small number of ray
trajectories, on the impedance. To isolate individual trajectories,
microwave absorbers are employed to cover some of the cavity walls,
and the experiment systematically includes short ray trajectories
involving bounces from exposed walls of the $1/4$-bow-tie cavity
(see insets of Fig. \ref{fig:walls_case} for examples), with no
perturbers present.

Fig. \ref{fig:walls_case} shows comparisons between the measured
impedance ($Z_{W}$) and the theoretical form ($Z^{(L_{M})}_{W}$).
Here the impedance $Z_{W}$ is measured from the microwave cavity
with specific walls ($W$) exposed, where $W=B$, $CD$, or $BC$ stands
for one or more of the walls A, B, C, and D shown in the insets of
Fig. \ref{fig:walls_case}. We have examined different combinations
of exposed walls and choose some representative cases to show here.
Figs. \ref{fig:walls_case} (a) and (b) are for the cases of one-port
experiments with (a) one wall exposed (wall B) and (b) two walls
exposed (walls C and D), and Figs. \ref{fig:walls_case} (c) and (d)
are for cases of two-port experiments with (c) one wall exposed
(wall B) and (d) two walls exposed (walls B and C). Notice that
$Z_{avg}^{(L_{M})}\rightarrow Z^{(L_{M})}_{W}$ in this case because
there is only one single realization. Thus, there is no ensemble
averaging and $p_{b(n,m)}=1$ for all trajectories in Eqs.
(\ref{eq:real part SOC}) and (\ref{eq:imaginary part SOC}).

Here we focus on the effects of ray trajectories, so we remove the
effect of the port mismatch from the measured impedance. We term
this quantity the impedance correction
\begin{equation}\label{eq:impedance_correction}
    z_{cor}\equiv R_{R}^{-1/2}(Z_{W}-Z_{R})R_{R}^{-1/2},
\end{equation}
and in Fig. \ref{fig:walls_case} the measured data (red and solid)
are constructed as the resistance correction $\textrm{Re}[z_{cor}]$
and the reactance correction $\textrm{Im}[z_{cor}]$. In each case,
the radiation impedance of the antenna ($Z_{R}$) is determined by a
separate measurement in which all four walls are covered by the
microwave absorbers \cite{S1}. According to Eq.
(\ref{eq:average_impedance}), the corresponding theoretical term is
\begin{equation*}\label{eq:theoretical_impedance_correction}
    z_{cor}^{(L_{M})}\equiv R_{R}^{-1/2}(Z^{(L_{M})}_{W}-Z_{R})R_{R}^{-1/2}
\end{equation*}
\begin{equation}\label{eq:theoretical_impedance_correction}
    \ \ \ \ \ \ =\rho^{(L_{M})}_{W}+i\chi^{(L_{M})}_{W}=z^{(L_{M})}_{W}.
\end{equation}
Therefore, the theoretical curves (blue dashed) of the resistance
correction and the reactance correction in Fig. \ref{fig:walls_case}
are $\textrm{Re}[z_{cor}^{(L_{M})}]$ and
$\textrm{Im}[z_{cor}^{(L_{M})}]$, respectively, and they can be
calculated by using Eq. (\ref{eq:SOC}) with the known geometry of
the cavity and port locations, including all short orbits up to the
maximum length $L_{M} = 200$ cm.

The measured data generally follow the theoretical predictions quite
well, thus verifying that the theory offers a quantitative
prediction of short-trajectory features of the impedance $Z_{W}$. In
Figs. \ref{fig:walls_case} (a) and (b), the one-port cases, we
examine the resistance (reactance) correction for the waves entering
and returning from the cavity through the single port, and the
short-trajectory corrections respectively include one trajectory for
$W=B$ and a sum over nine trajectories for $W=CD$. In Figs.
\ref{fig:walls_case} (c) and (d), the two-port cases, we examine the
resistance (reactance) correction  between the two ports. It
corresponds to the elements $z_{cor,1,2}$ and
$z_{cor,1,2}^{(L_{M})}$ in the $2\times 2$ matrices ($z_{cor}$ and
$z_{cor}^{(L_{M})}$), and the short-trajectory corrections
respectively include sums over two trajectories $(W=B)$ and four
trajectories $(W=BC)$. For illustration, the insets of Fig.
\ref{fig:walls_case} show some representative short ray
trajectories. Note the direct trajectory between the two ports
without bouncing on walls is treated as a ray trajectory in Eqs.
(\ref{eq:real part SOC}) and (\ref{eq:imaginary part SOC}).

\begin{figure}
\includegraphics[height=1.8in,width=2.4in]{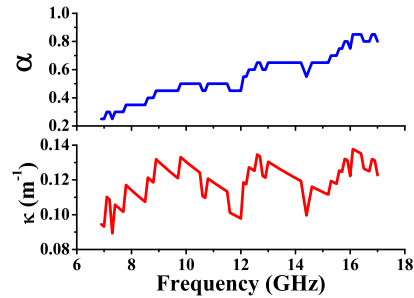}
\caption{(Color online) Plot of the loss parameter $\alpha$ and the
attenuation parameter $\kappa$ versus frequency inside the empty
$1/4$-bow-tie cavity.} \label{fig:loss}
\end{figure}

For the propagation attenuation, a frequency-dependent attenuation
parameter $\kappa(f)$ is calculated utilizing the previously
measured frequency-dependent loss parameter $\alpha$
\cite{SamThesis} for this cavity. Fig. \ref{fig:loss} shows the loss
parameter $\alpha(f)$ and the attenuation parameter $\kappa(f)$ for
the empty cavity case. Here $\alpha$ is defined as the ratio of the
closed-cavity mode resonance 3-dB bandwidth to the mean spacing
between cavity modes, $\alpha = k^{2}/(\Delta k^{2}Q) \simeq
k^{2}A/(4\pi Q)$, where $A$ is the area of the cavity
($A=0.115m^{2}$), $k$ is the wavenumber, and $Q$ is the quality
factor of the cavity. The mean spacing between modes varies from 21
MHz at 6 GHz to 6.9 MHz at 18 GHz. We obtain the loss parameter
$\alpha$ from the measured impedance data in the empty cavity case
according to the procedures presented in previous work \cite{S1, S3,
SamThesis}. In this procedure the frequency-dependent loss parameter
$\alpha(f)$ is determined by selecting a proper frequency range
(e.g., 1.8 GHz), computing the PDF of the ``perfectly coupled''
normalized impedance $i\xi_{0}$ (Eq.
(\ref{eq:normalized_impedance})), and then comparing to different
PDFs that were generated from numerical RMT, using $\alpha$ as a
fitting parameter. Once $\alpha(f)$ is known, the
frequency-dependent attenuation parameter $\kappa(f)$ can be
calculated because the dominant attenuation comes from losses in the
top and bottom plates of the cavity, and it is well modeled by
assuming that the waves suffer a spatially uniform propagation loss
$\kappa=k/(2Q)=2\pi\alpha/(kA)$.

\subsection{Sources of errors}\label{subsec:errors}
\begin{figure}
\includegraphics[height=1.8in,width=2.4in]{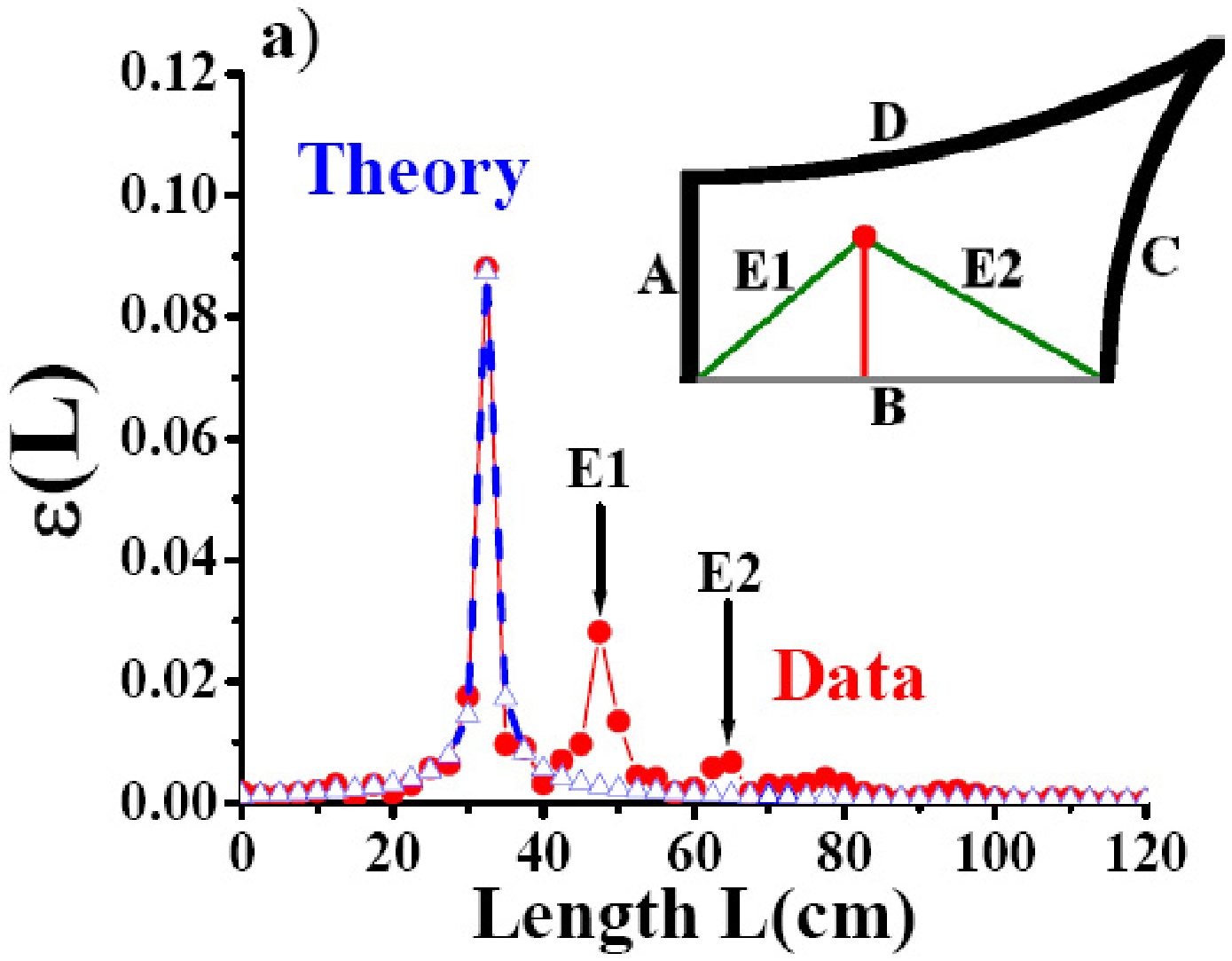}
\includegraphics[height=1.8in,width=2.4in]{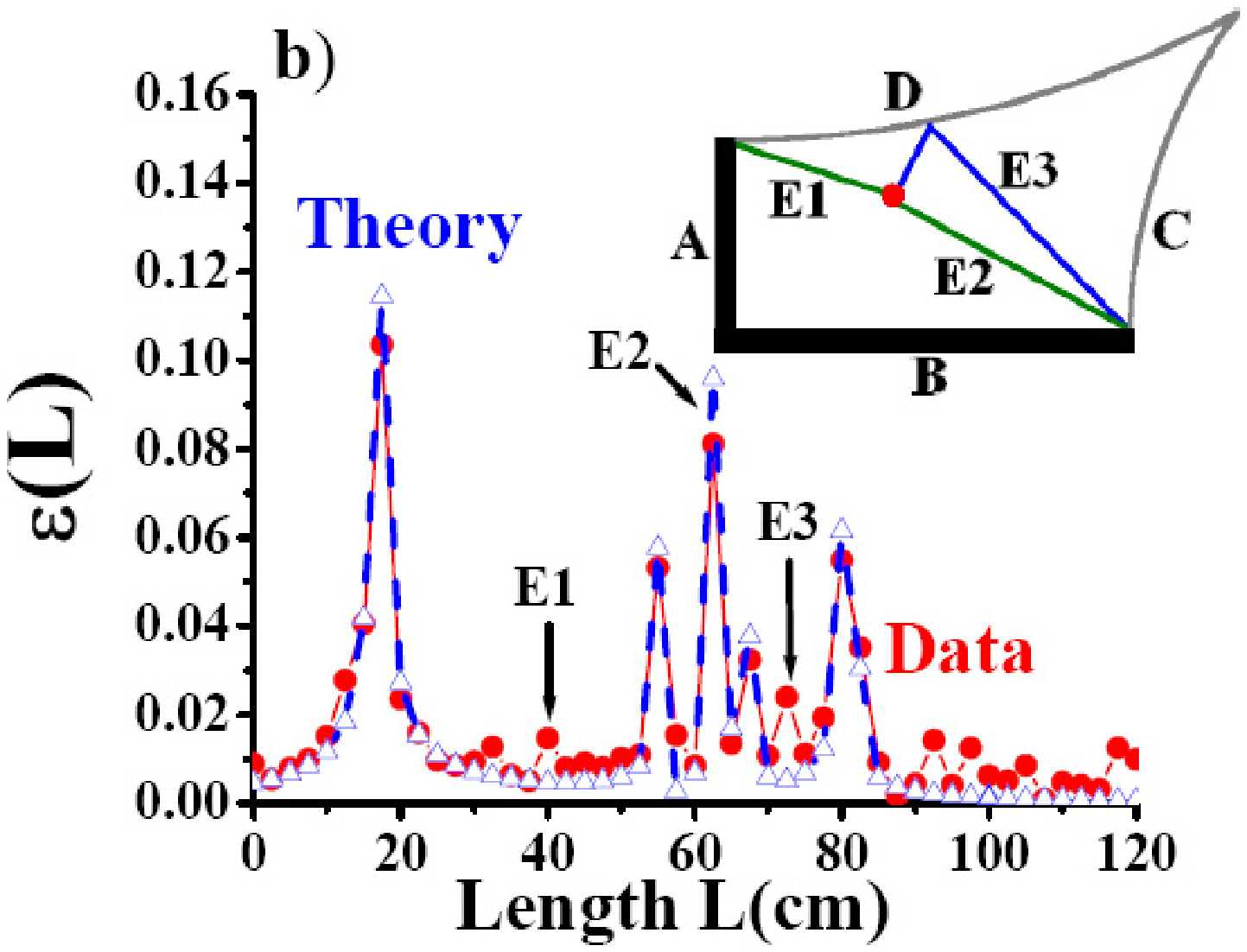}
\includegraphics[height=1.8in,width=2.4in]{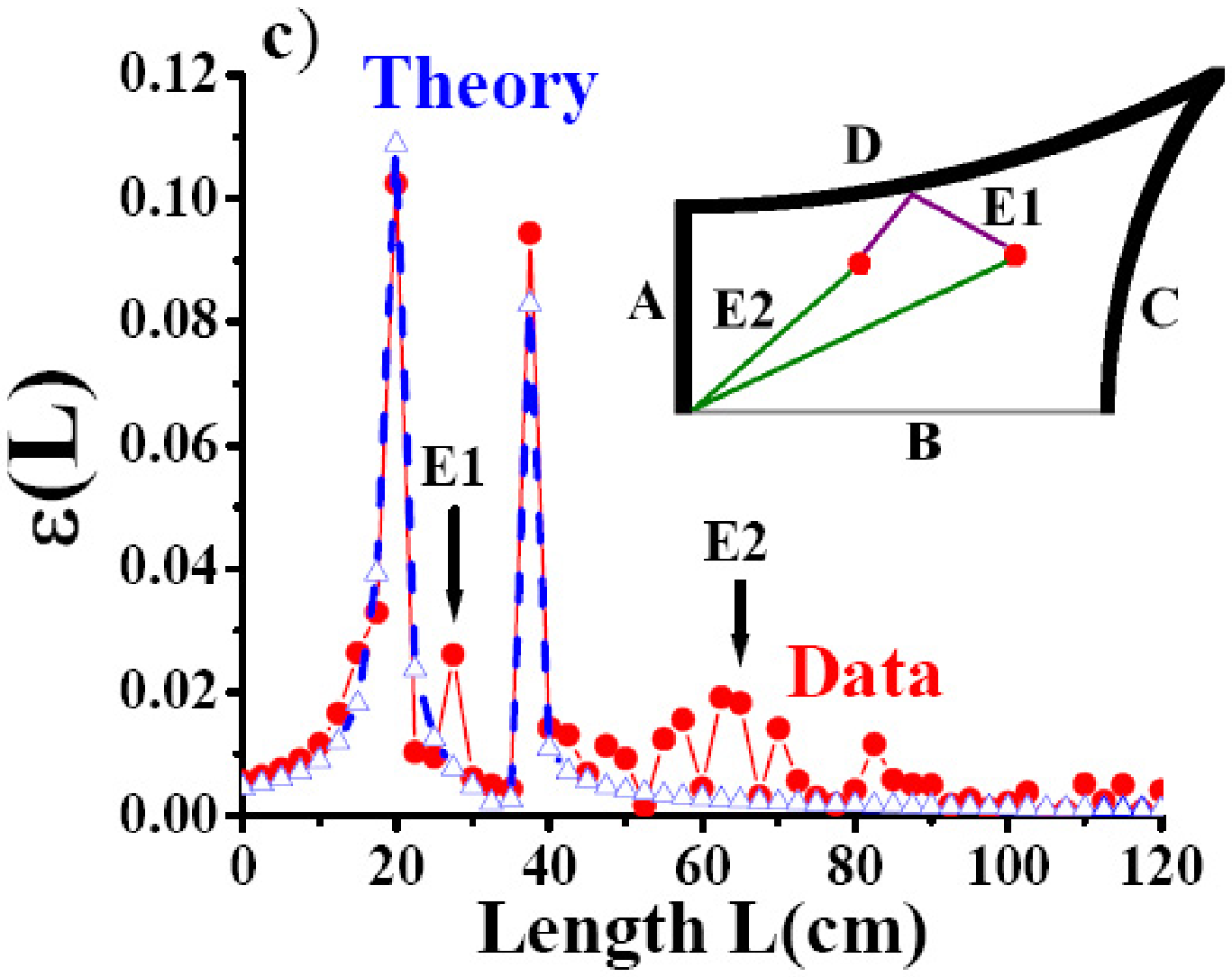}
\includegraphics[height=1.8in,width=2.4in]{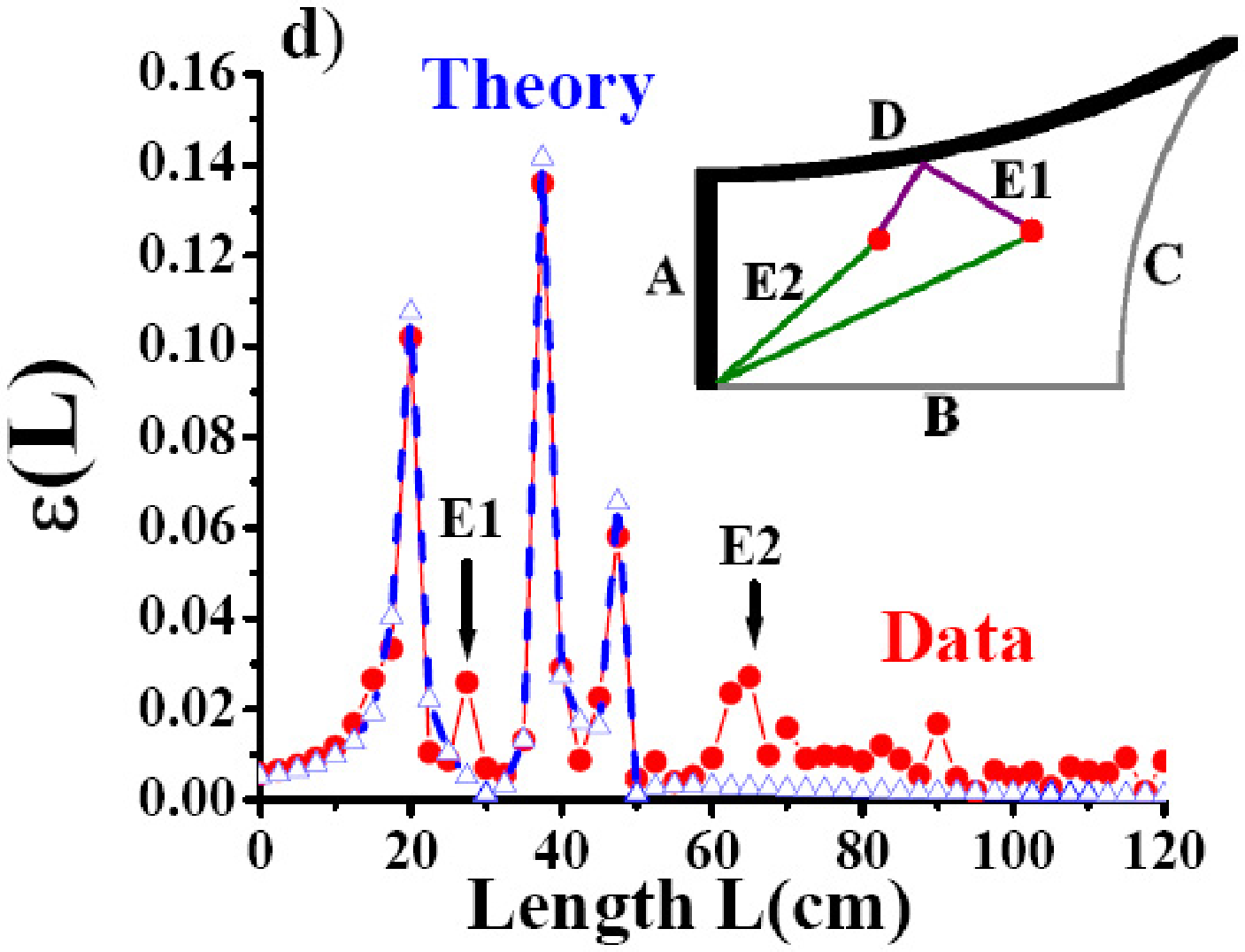}
\caption{(Color online) Plot of the absolute values of the impedance
correction in the length domain $\varepsilon(L)$ in the cases of (a)
one-port with one wall (wall B), (b) one-port with two walls (walls
C and D), (c) two-port with one wall (wall B), and (d) two-port with
two walls (walls B and C) short trajectories, versus length from 0
to 120 cm inside the $1/4$-bow-tie cavity. Shown are the measured
data in red curves with circular symbols and the theory in blue
dashed curves with triangular symbols. The insets show the
$1/4$-bow-tie billiard with labeled trajectories that show the
sources of error (E1, E2, and E3).} \label{fig:LDomain}
\end{figure}

We propose that there are two major sources of the deviations
between the theory and experiment shown in Fig.
\ref{fig:walls_case}. The first is that the microwave absorbers do
not fully suppress the trajectories, and the second arises from the
ends of microwave absorbers that scatter energy back to the ports.
To verify this, the $k=\omega/c$ dependent impedance corrections
data in Fig. \ref{fig:walls_case} are Fourier transformed to the
length domain and shown in Fig. \ref{fig:LDomain} as
$\varepsilon(L)_{meas}\equiv c|\textrm{FT}\{z_{cor}\}|$ and
$\varepsilon(L)_{theory}\equiv c|\textrm{FT}\{z_{cor}^{(L_{M})}\}|$,
where $c$ is the speed of light, and $\textrm{FT}\{...\}$ is the
Fourier transformation ($k\rightarrow L$). Note that the frequency
range of the Fourier transformation is from 6 GHz to 18 GHz, and
therefore, the resolution in length is 2.5 cm.

In the length domain, the major peaks of the measured data (red)
match the peaks of the theoretical prediction (blue and dashed), and
this verifies that the short-trajectory correction can describe the
major features of the measured impedance in the scattering system.
For example, the matched peak in the theory curve and the data curve
in Fig. \ref{fig:LDomain}(a) corresponds to the short trajectory
from the port to wall B and returning, shown as the red (vertical)
line in the inset of \ref{fig:LDomain}(a). However, there are
several minor peaks in the measured data not present in the
theoretical curves. After further examination of the geometry, the
positions of these deviations in Fig. \ref{fig:LDomain} match the
lengths of trajectories which are related to the partially-blocked
corners of the cavity, or bounce off microwave absorbers with a
large incident angle. When the microwave absorbers end at the
corners, they produce gaps and edges, and these defects create weak
diffractive short trajectories. For example, the green lines E1 and
E2 in the insets in Figs. \ref{fig:LDomain}(a) and (b) and E2 in
Figs. \ref{fig:LDomain}(c) and (d) represent the diffractive short
trajectories leaving a port, bouncing off the partially covered
corners, and returning to a port. Their path lengths match the
deviations between the measured data and the theory as labeled in
the figure. Furthermore, the blue line E3 in the inset in Fig.
\ref{fig:LDomain}(b) represents the short trajectories produced by
the corner and bounced from one wall.

The other source of error is imperfection of the microwave absorbers
that reflect $\sim-20$ dB of the incident signal for normal
incidence, and more for oblique incidence. Therefore the short
trajectories shown as purple lines E1 in Figs. \ref{fig:LDomain}(c)
and (d) with a large incident angle bring about deviations between
the theory and experimental data. These sources of error due to the
ends of absorbers or large incident angles on absorbers were not
included in the short-trajectory correction. However, these errors
will not concern us further because all microwave absorbers are
removed from the cavity in the experiments discussed below.

Besides the errors discussed above, another source of error is the
difficulty in reproducing the antenna geometry with each measurement
as the cavity is opened and re-sealed between the measurement of the
radiation impedance and exposed wall cases. Another concern is
multiply-reflected trajectories that bounce off of the antennas.
However, because we describe trajectories in terms of the impedance
instead of the scattering parameter, the multiply-reflected
trajectories are incorporated in a single impedance term (see
\cite{impedance}). This is an important advantage of using impedance
because it can take account of the multiple-reflected trajectories
in a simple compact form.

\subsection{Short-trajectory correction in the empty cavity}\label{subsec:empty_cavity}
\begin{figure}
\includegraphics[height=2.0in,width=2.4in]{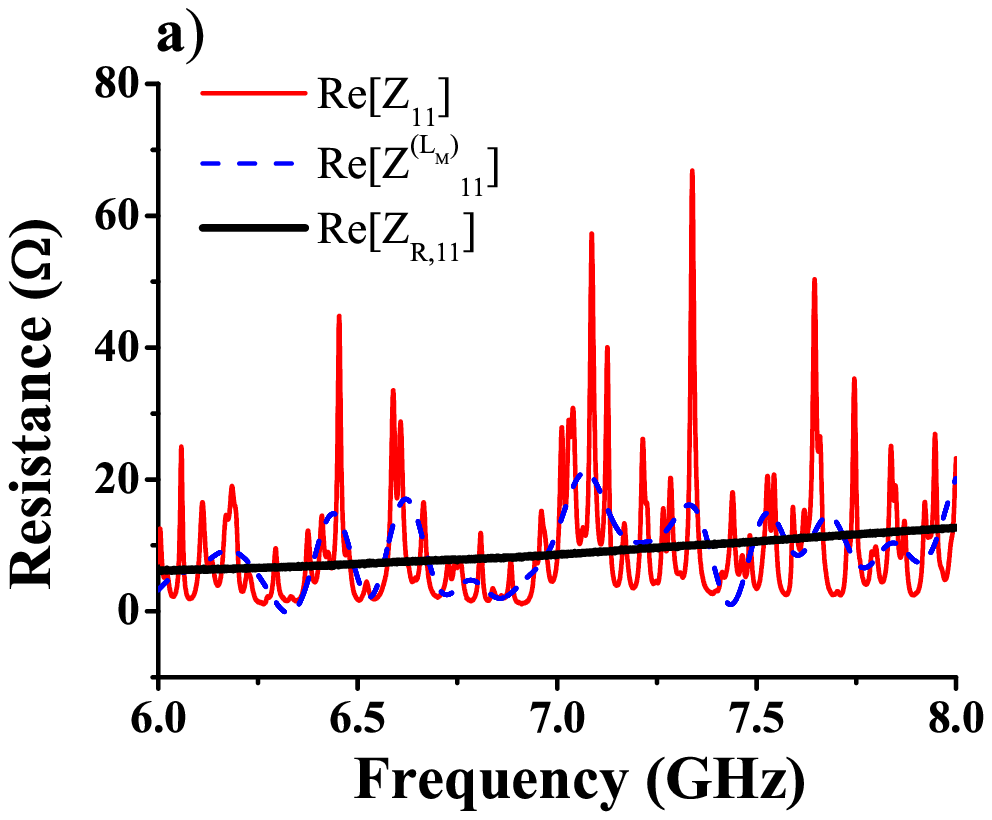}
\includegraphics[height=2.0in,width=2.4in]{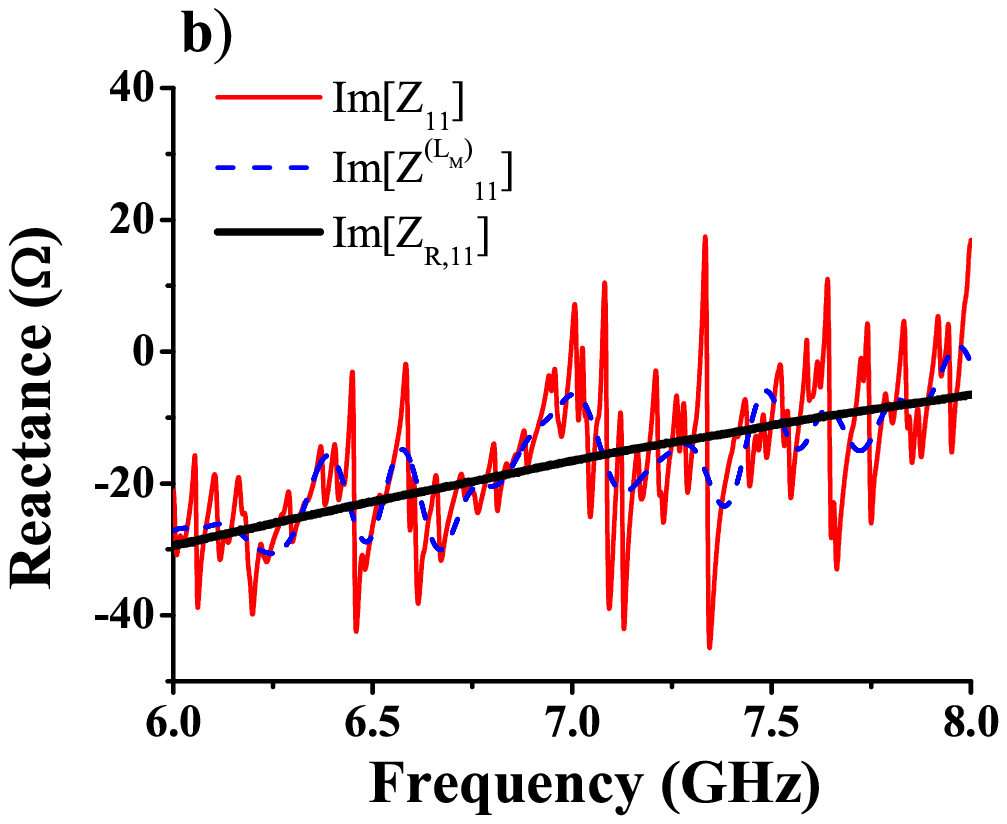}
\includegraphics[height=2.0in,width=2.4in]{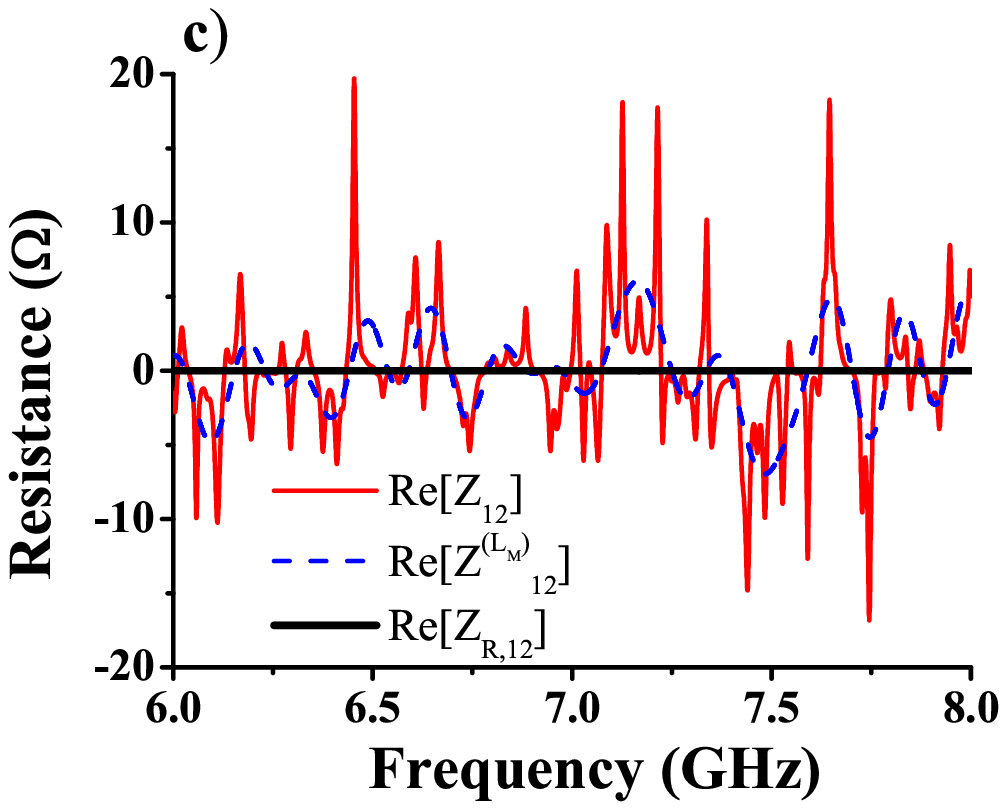}
\includegraphics[height=2.0in,width=2.4in]{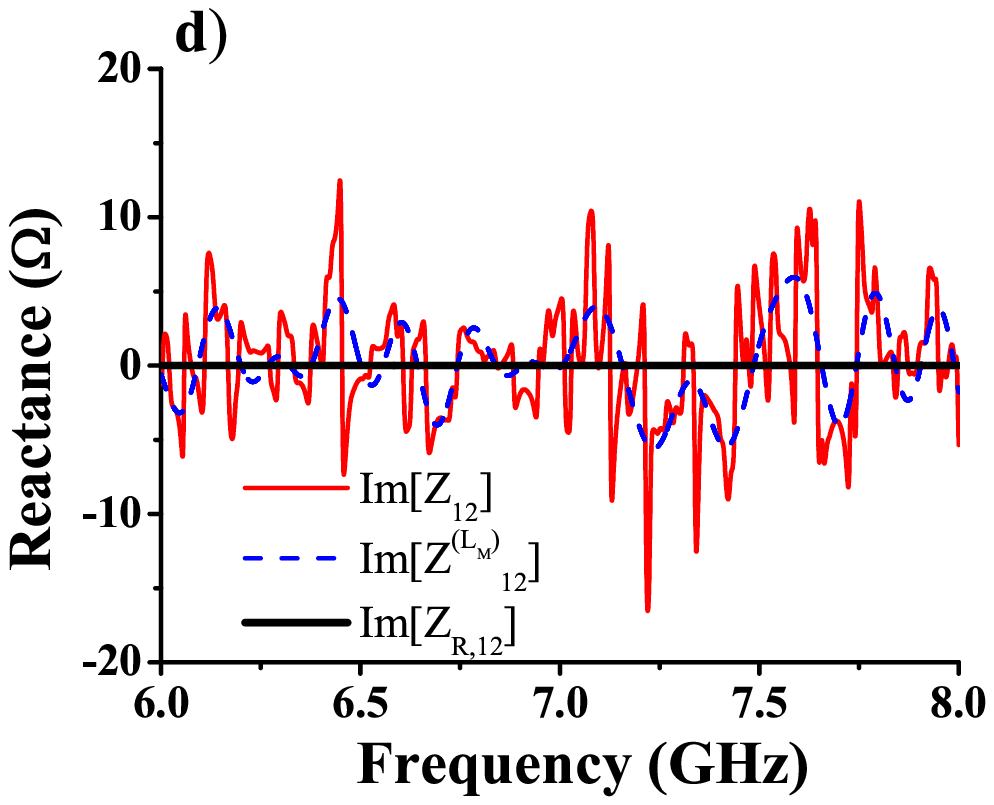}
\caption{(Color online) Plot of the impedance due to the
trajectories of four side-walls, versus frequency from 6 to 8 GHz
inside the empty $1/4$-bow-tie cavity. Shown are (a) the real part
of $Z_{11}$, (b) the imaginary part of $Z_{11}$, (c) the real part
of $Z_{12}$ and (d) the imaginary part of $Z_{12}$. The three curves
are the measured impedance $Z$ (red thinner), the theoretical
impedance $Z^{(L_{M})}$ (blue dashed), as well as the radiation
impedance $Z_{R}$ (black).} \label{fig:empty_impedance}
\end{figure}

    In the next experiment, an empty $1/4$-bow-tie cavity with no microwave
absorbers or perturbers is used. Therefore, all possible
trajectories between ports are present in this single realization.
Fig. \ref{fig:empty_impedance} shows (a) the real and (b) the
imaginary parts of the first diagonal component of the impedance in
a two-port cavity, corresponding to $\textrm{Re}[Z_{11}]$ and
$\textrm{Im}[Z_{11}]$. Figs. \ref{fig:empty_impedance} (c) and (d)
are for the off-diagonal component of the impedance,
$\textrm{Re}[Z_{12}]$ and $\textrm{Im}[Z_{12}]$. The radiation
impedance (black) traces through the center of the fluctuating
impedance data of the empty cavity and represents the nonuniversal
aspects of the coupling antennas
\cite{Henry_paper_one_port,Henry_paper_many_port, S1}. Note for the
$Z_{12}$ case, all signals from port 1 to port 2 are treated as
trajectories, so the off-diagonal radiation impedance is zero. Also
shown in Fig. \ref{fig:empty_impedance}, the theoretical impedance
$Z^{(L_{M})}$ (Eqs. (\ref{eq:real part SOC}) and (\ref{eq:imaginary
part SOC})) include a finite number of short trajectories
($L_{M}=200$ cm, given a total of 584 trajectories for
$Z^{(L_{M})}_{11}$ and 1088 trajectories for $Z^{(L_{M})}_{12}$).
Note that the attenuation parameter $\kappa(f)$ is determined
through the same procedures as in Sec. \ref{subsec:wall_cases}.

In Fig. \ref{fig:empty_impedance}, the theoretical impedance
$Z^{(L_{M})}$ tracks the main features of the single-realization
measured impedance although there are many sharp deviations between
the two sets of curves. These fluctuations are expected because of
the infinite number of trajectories ($L_{b(n,m)}>200$ cm) not
included in the theory.

\begin{figure}
\includegraphics[scale=0.7]{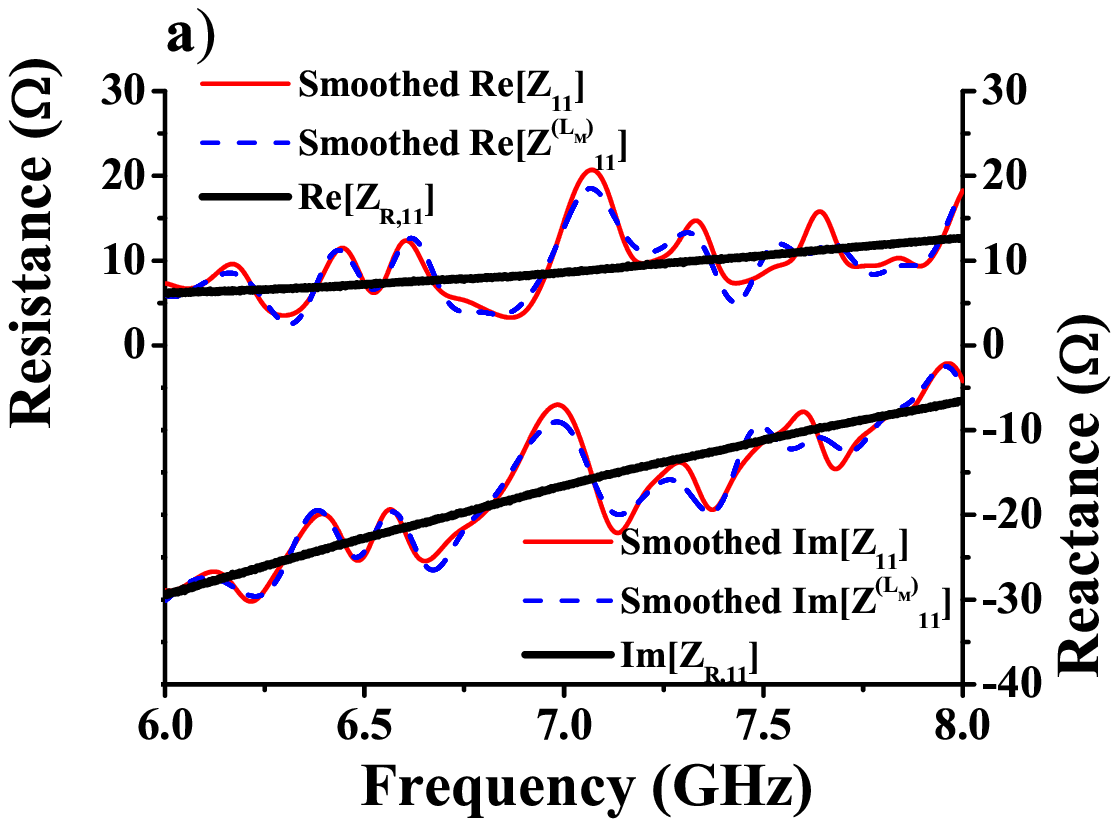}
\includegraphics[scale=0.7]{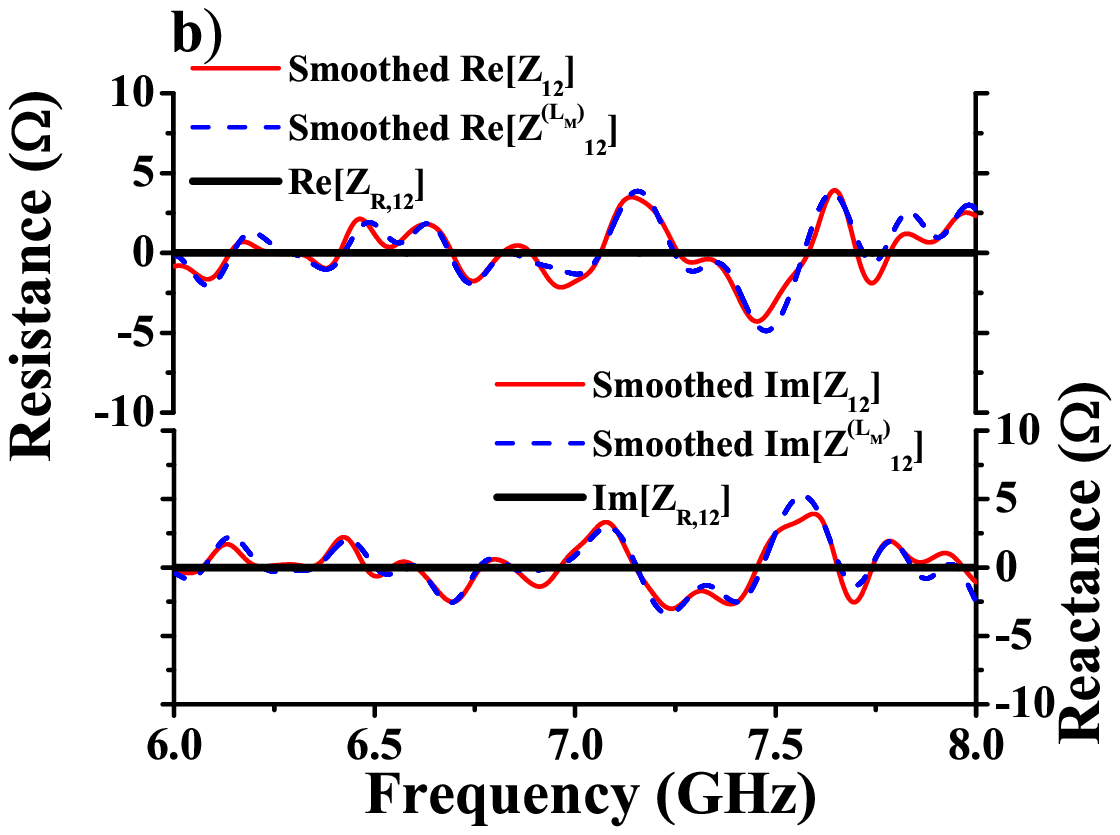}
\caption{(Color online) Plot of the smoothed impedance versus
frequency from 6 to 8 GHz. Shown are, (a) for $Z_{11}$ and (b) for
$Z_{12}$, the real (three upper curves) and the imaginary part
(three lower curves) of the smoothed impedance for the theory
($Z^{(L_{M})}$ with $L_{M}=200$ cm, blue dashed) and the experiment
(red solid), as well as the measured (un-smoothed) radiation
impedance of the port ($Z_{R}$, black thick).}
\label{fig:smoothed_impedance}
\end{figure}

It is more appropriate to compare the theory to a frequency ensemble
of single-realization data. A frequency ensemble is created by
considering frequency smoothed experimental data and comparing it
with the smoothed theoretical prediction for system-specific
contributions to the impedance. The frequency smoothing (Eqs.
(\ref{eq:Gaussian_smoothing}) and (\ref{eq:Gaussian_function}))
suppresses the impedance fluctuations due to long trajectories and
reveals the features associated with short trajectories. Fig.
\ref{fig:smoothed_impedance} shows the radiation impedance (black
thick), the smoothed measured impedance $Z$ (red solid) and the
smoothed theoretical impedance $Z^{(L_{M})}$ (blue dashed). The
smoothing is made by a Gaussian smoothing function with the standard
deviation $\Delta\omega/(2\pi)=240$ MHz (Eq.
(\ref{eq:Gaussian_function})). Gaussian frequency smoothing inserts
an effective low-pass Gaussian filter on the trajectory length, and
thus, the components of impedances ($Z$ and $Z^{(L_{M})}$) in the
length domain are limited to those with the path length $L\lesssim
2\pi c/\Delta\omega=125$ cm. Fig. \ref{fig:smoothed_impedance} shows
that the smoothed theory matches the similarly smoothed experimental
data very well, therefore, the short trajectory theory correctly
captures the effects of ray trajectories out to this length region
($\lesssim125$ cm).

When computing the sum of short-trajectory correction terms in a low
loss case like the empty cavity, a problem appears associated with
the finite number of terms ($L_{b(n,m)}\leq L_{M}$) in the sum over
trajectories. In the theory, to perfectly reproduce the measured
data in the empty cavity requires an infinite number of correction
terms.  Therefore, in some frequency regions where the experimental
impedance changes rapidly, the finite sum for the theoretical
resistance $\textrm{Re}[Z^{(L_{M})}]$ will show values less than
zero, which are not physical for a passive system. For example, see
the blue dashed curve in Fig. \ref{fig:empty_impedance}(a) between
7.3 and 7.4 GHz. This problem is similar to Gibbs phenomenon in
which the sum of a finite number of terms of the Fourier series has
large overshoots near a jump discontinuity.

\begin{figure}
\includegraphics[scale=0.6]{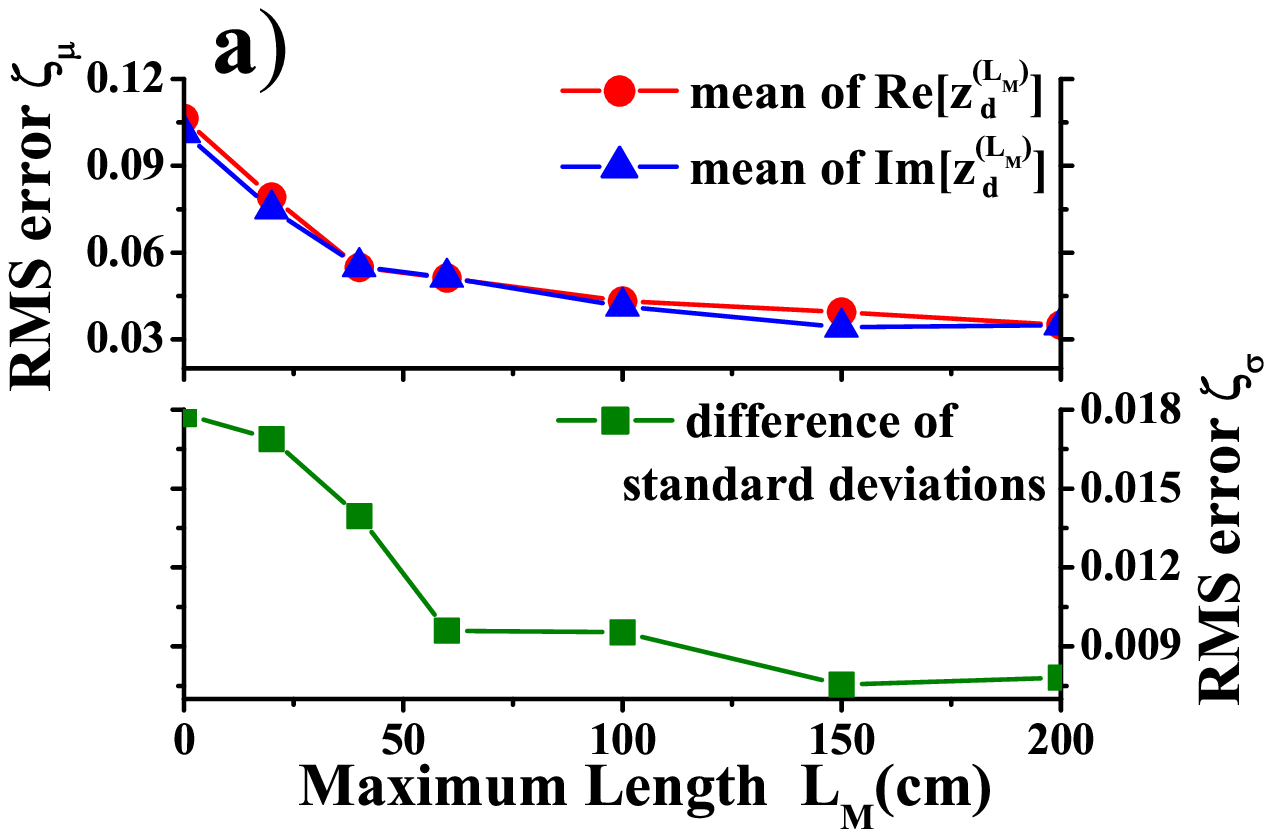}
\includegraphics[scale=0.6]{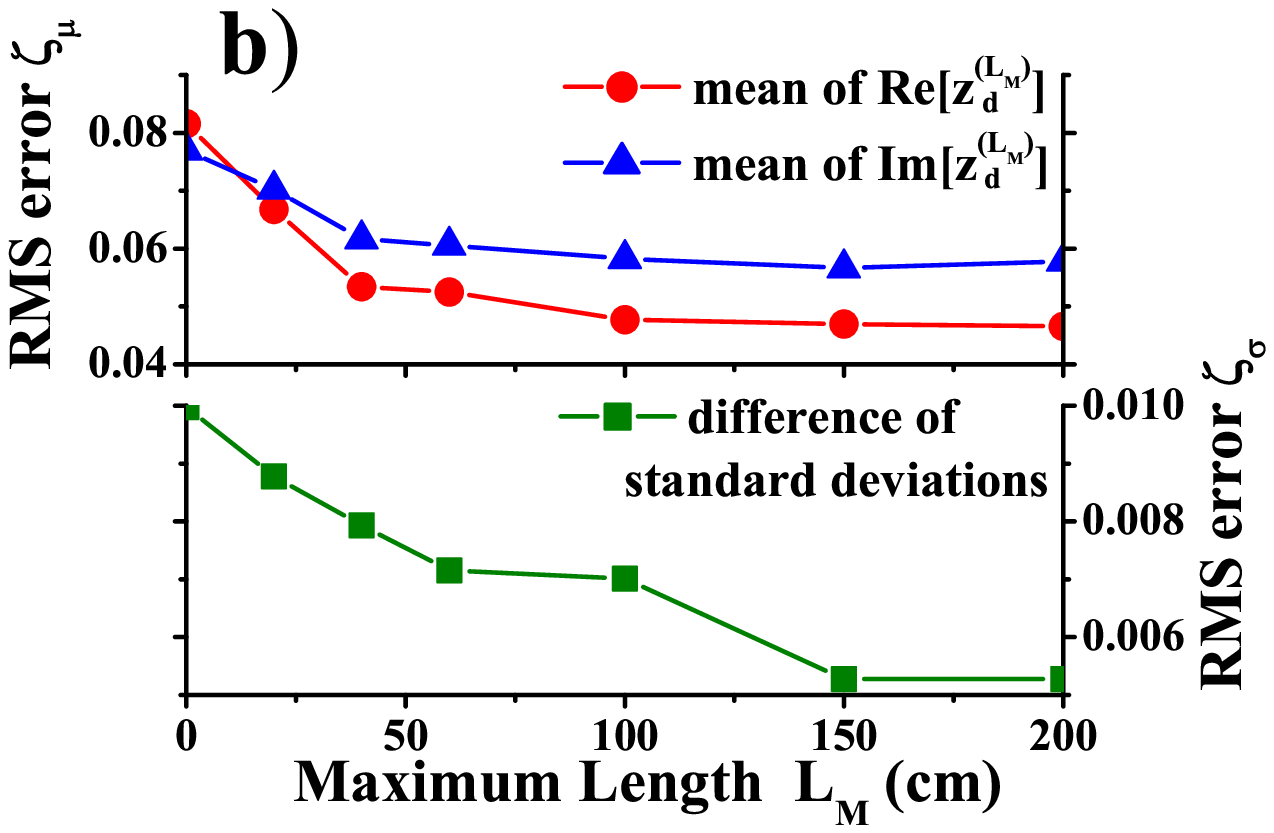}
\caption{(Color online) Plot of the RMS errors for (a) the one-port
experiment and (b) the two-port experiment. Shown are $\zeta_{\mu}$
for the mean of the real part (red circles) and the imaginary part
(blue triangles) of the normalized impedance difference and
$\zeta_{\sigma}$ for the difference of standard deviations (green
rectangles), versus short-trajectory corrections with the maximum
length from 0 to 200 cm. Data are taken for a single realization of
the bow-tie cavity.} \label{fig:PDF_impedance}
\end{figure}

Besides verifying that the short-trajectory correction agrees with
system-specific features of the measured data, we would next like to
demonstrate that including short-trajectory corrections improves the
ability to reveal underlying universal statistical properties, even
in a single realization of the system. We compute the statistical
properties of the real and imaginary part of the normalized
impedance difference
\begin{equation}\label{eq:impedance_difference}
    z^{(L_{M})}_{d}\equiv R_{R}^{-1/2}(Z-Z^{(L_{M})})R_{R}^{-1/2}
\end{equation}
in 500 MHz frequency windows from 6 to 18 GHz for a single
realization of the bow-tie cavity. Random matrix theory predicts
that the distribution of $\textrm{Re}[z^{(L_{M})}_{d}]$ and
$\textrm{Im}[z^{(L_{M})}_{d}]$ should have zero means, and identical
standard deviations \cite{Fyodorov, S1,S2,S3, henrythesis,
Experimental_tests_our_work}. For the one-port case, we use the
normalized impedance difference value directly, and for the two-port
case, we consider the eigenvalues of the $2\times2$ normalized
impedance difference matrix.

Fig. \ref{fig:PDF_impedance} shows that the RMS errors of
statistical parameters between the measured data and the theory
decrease upon including more short trajectories in the correction
(i.e., increasing $L_{M}$). We compute the root mean square value of
errors ($\zeta$ defined later) for a series of frequency windows
covering the range from 6 to 18 GHz, and the results are shown
versus different short-trajectory corrections with varied maximum
lengths $L_{M}$. Notice the cases of $L_{M}=0$ denote the impedance
corrected by only the radiation impedance without any
short-trajectory correction. For the normalized impedance difference
$z_{d}^{(L_{M})}$ in each frequency window, the RMS error
$\zeta_{\mu}$ is defined as the root mean square value of $|\mu_{R}
- 0|$ or $|\mu_{X} - 0|$ for the difference of the measured mean
from zero, and $\zeta_{\sigma}$ is defined as the root mean square
value of $|\sigma_{R}-\sigma_{X}|/(\sigma_{R}+\sigma_{X})$ for the
difference of standard deviations between the real part PDF and the
imaginary part PDF. Here $\mu_{R}$ and $\mu_{X}$ are respectively
the means of $\textrm{Re}[z^{(L_{M})}_{d}]$ and
$\textrm{Im}[z^{(L_{M})}_{d}]$ in each window, and $\sigma_{R}$ and
$\sigma_{X}$ are the standard deviations of
$\textrm{Re}[z^{(L_{M})}_{d}]$ and $\textrm{Im}[z^{(L_{M})}_{d}]$ in
each window. In all three statistical parameters for both one-port
and two-port cases, the RMS errors $\zeta$ decrease when we correct
the data with more short trajectories. This verifies that using
short-trajectory corrections in a single realization of the
wave-chaotic system can more effectively reveal the universal
statistical properties in the data.

\subsection{Configuration averaging approach}\label{subsec:ensemble}

Many efforts to determine universal RMT statistics in experimental
systems are based on a configuration averaging approach that creates
an ensemble average from realizations with varied configurations. In
principle, one can recover the nonuniversal properties of the system
\cite{Brouwer_Lorentzian, Kuhl} via the configuration averaging
approach, which is motivated by the ``Poisson kernel'' theory of
Mello, Pereyra, and Seligman \cite{Poisson_Kernel_Original}.
Specifically, ensemble averages of the measured cavity data are used
to remove the system-specific features in each single realization.
Note that in the past, the configuration averaging approach was
explicitly assumed to only remove the effects of the nonuniversal
coupling; however, it was recently generalized to include the
nonuniversal contributions of short trajectories
\cite{Bulgakov_Gopar_Mello_Rotter}.

\begin{figure}
\includegraphics[height=2.4in,width=3.2in]{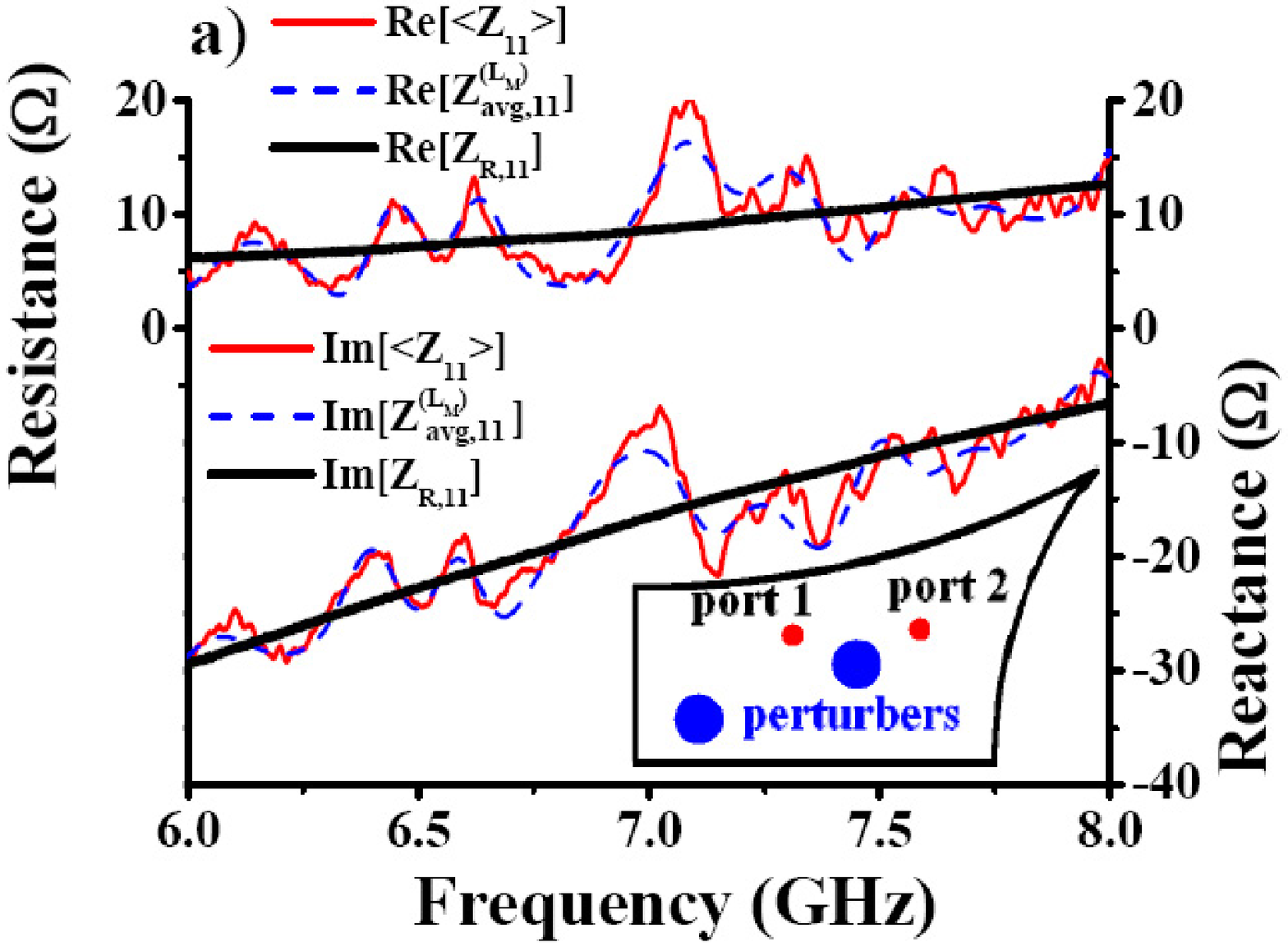}
\includegraphics[height=2.4in,width=3.2in]{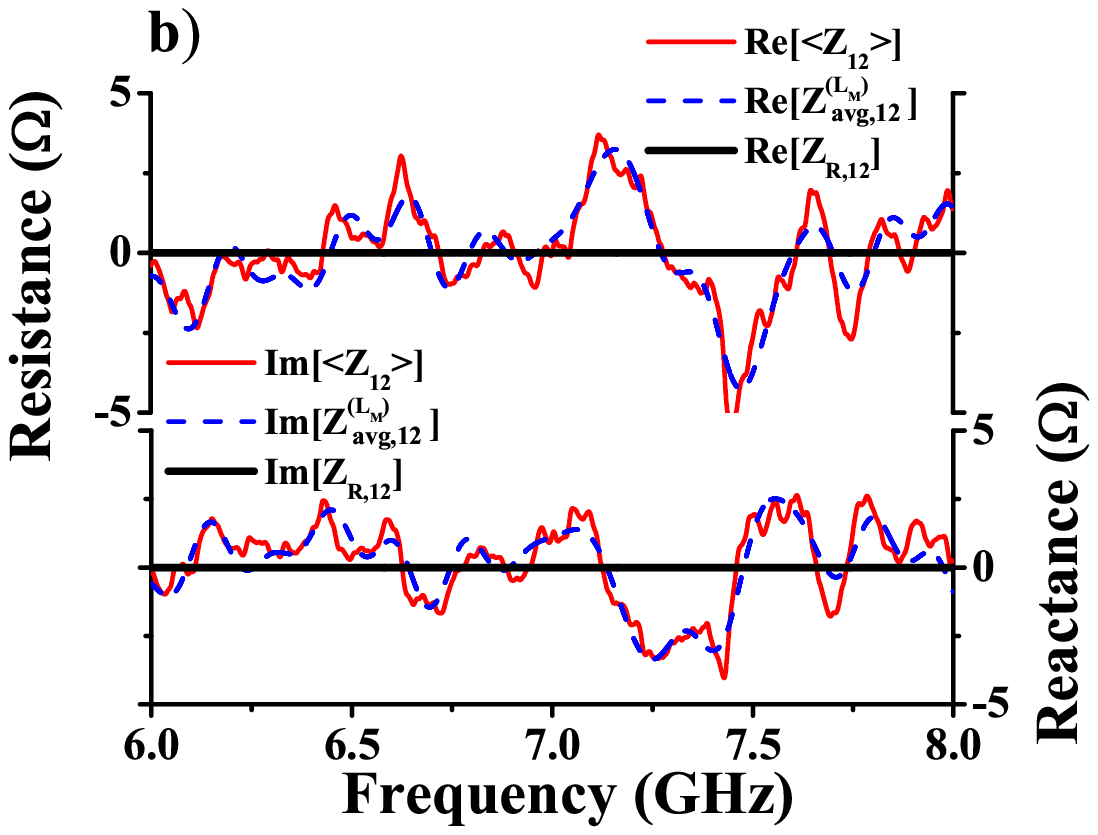}
\includegraphics[height=2.4in,width=3.0in]{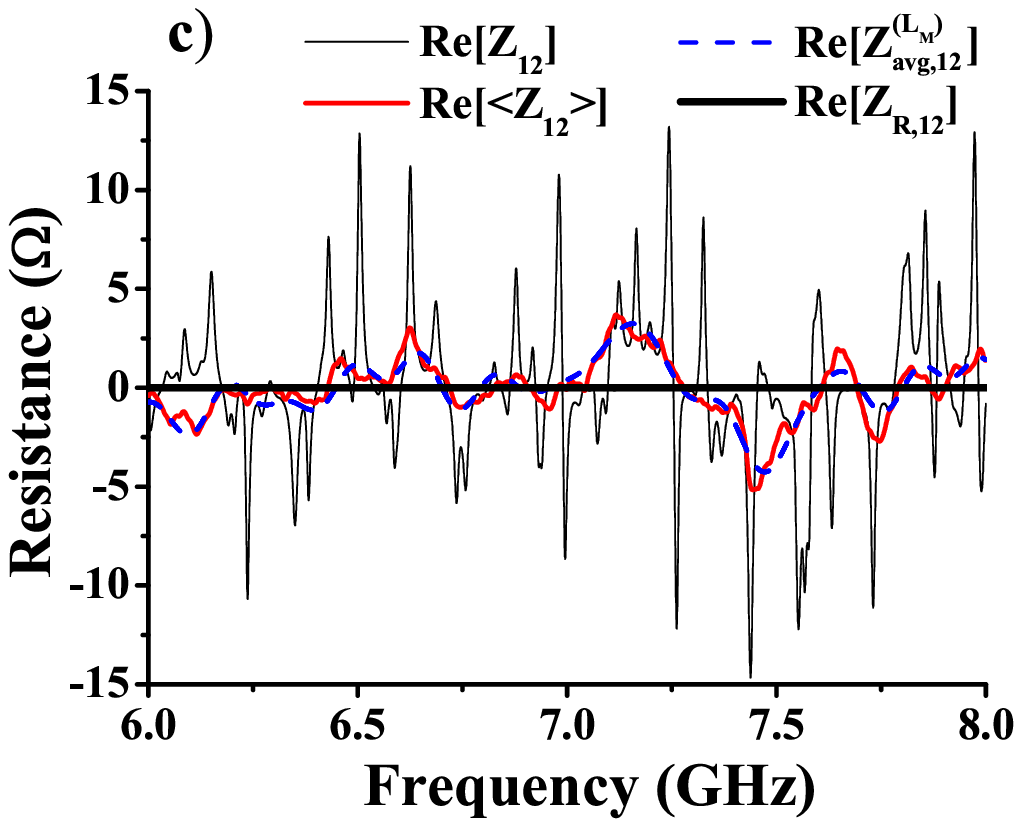}
\caption{(Color online) Plot of the average impedance versus
frequency from 6 to 8 GHz. Shown are, (a) for $Z_{11}$ and (b) for
$Z_{12}$, the real (three upper curves) and the imaginary part
(three lower curves) of the average impedance for the theory
($Z^{(L_{M})}_{avg}$ with $L_{M}=200$ cm, blue dashed) and the
configuration average experiment ($\langle Z\rangle$ red solid), as
well as the measured radiation impedance of the ports ($Z_{R}$,
black thick). Plot (c) shows the real part of the measured impedance
$Z_{12}$ (black thin) in a single realization comparing with
$Z^{(L_{M})}_{avg,12}$ and $\langle Z_{12}\rangle$. Inset: The
wave-chaotic two-dimensional cavity with perturbers and two ports.}
\label{fig:ensemble_impedance}
\end{figure}

Here the experimental results verify that the short-trajectory
correction (Eqs. (\ref{eq:real part SOC}) and (\ref{eq:imaginary
part SOC})) can describe nonuniversal characteristics of
wave-chaotic systems in the configuration ensemble. Two cylindrical
pieces of metal are added as perturbers in the wave-chaotic system
that is shown in the inset of Fig. \ref{fig:ensemble_impedance},
where the dots represent the ports and the two circles represent the
perturbers. The locations of the two perturbers inside the cavity
are systematically changed and accurately recorded to produce a set
of 100 realizations for the ensemble
\cite{S2,S3,Experimental_tests_our_work}. The scattering matrix $S$
is measured from 6 to 18 GHz, covering roughly 1070 modes of the
closed cavity. After the ensemble average, longer ray trajectories
have higher probability of being blocked by the two perturbers in
the 100 realizations; therefore, the main nonuniversal contributions
are due to shorter ray trajectories. We compare the measured
ensemble averaged impedance $\langle Z\rangle$ and the theoretical
impedance $Z_{avg}^{(L_{M})}$ that is calculated from Eqs.
(\ref{eq:real part SOC}) and (\ref{eq:imaginary part SOC}) with the
maximum trajectory length $L_{M}=200$ cm.

In the configuration ensemble, contrary to the previous cases
without perturbers, we need to introduce a survival probability
$p_{b(n,m)}$ for each ray trajectory term in Eqs. (\ref{eq:real part
SOC}) and (\ref{eq:imaginary part SOC}). Notice that the two
perturbers can block ray trajectories and influence their presence
in the ensemble realizations. Thus, we multiply each term in the sum
by a weight $p_{b(n,m)}$ equal to the fraction of perturbation
configurations in which the trajectory is not intercepted by the
perturbers. The values of $p_{b(n,m)}$ are between 0 and 1, and a
longer ray trajectory generally has a higher chance of being blocked
by perturbers, and thus it has smaller $p_{b(n,m)}$.

Note that by recording the positions of perturbers in all
realizations, we are able to do impedance normalization with
short-trajectory correction individually for each realization,
similar to the procedures in the empty cavity case. Here we
introduce $p_{b(n,m)}$ as a more general description for the case in
which only the probabilities of survival of particular short
trajectories are known. In addition, we ignore the effect of newly
created trajectories by the perturbers in each specific realization
because they are averaged out in the ensemble. Note that the
attenuation parameter $\kappa$ in the short-trajectory correction
terms (Eqs. (\ref{eq:real part SOC}) and (\ref{eq:imaginary part
SOC})) is recalculated using the measured PDFs of impedance in the
ensemble case, using procedures similar to those for the case of the
empty cavity. Due to the presence of two perturbers in the cavity,
the attenuation parameter and loss parameter are slightly larger
($\sim 0.1$ for $\alpha$) than in the empty cavity case.

The result of comparisons between the $2\times2$ matrices $\langle
Z\rangle$ and $Z^{(L_{M})}_{avg}$ of the two-port experiment is
shown in Fig. \ref{fig:ensemble_impedance}. We have published the
result of the one-port experiment in another paper \cite{Jenhao}.
Here Fig. \ref{fig:ensemble_impedance} (a) shows the comparison of
the first diagonal component of the impedance, and Fig.
\ref{fig:ensemble_impedance} (b) shows the comparison of the
off-diagonal component. The measured data (red solid) follow the
trend of the radiation impedance (black thick), and the theory (blue
dashed) reproduces most of the fluctuations in the data by including
only a modest number of short-trajectory correction terms. Fig.
\ref{fig:ensemble_impedance}(c) illustrates the comparison between
the measured impedance $Z$ in a single realization and $\langle
Z\rangle$ The strong fluctuations in the measured impedance $Z$
(black thin) curve have diminished due to the configuration
averaging, and the remaining fluctuations of the configuration
averaged impedance $\langle Z\rangle$ away from the radiation
impedance are closely tracked by the theoretical (blue dashed)
curve. Note that no wavelength averaging is used here. The good
agreement between the measured data and the theoretical prediction
verifies that the new theory, Eqs. (\ref{eq:real part SOC}) and
(\ref{eq:imaginary part SOC}), predicts the nonuniversal features
embodied in the ensemble averaged impedance well.

The deviations between the $\langle Z\rangle$ curves and the
$Z^{(L_{M})}_{avg}$ curves in Fig. \ref{fig:ensemble_impedance} may
come from several effects. The first is the remaining fluctuations
in the $\langle Z\rangle$ due to the finite number of realizations.
We estimate this to be on the order of $\sigma_{\langle
Z\rangle}\simeq\sigma_{Z}/\sqrt{100}\sim1\Omega$, where
$\sigma_{\langle Z\rangle}$ is the standard deviation of ensemble
averaged impedance $\langle Z\rangle$ of 100 realizations, and
$\sigma_{Z}$ is the standard deviation of the measured impedance $Z$
of a single realization. This accounts for the remaining sharp
features in $\langle Z\rangle$. Another source of errors is that we
ignore the effect of newly created trajectories by the perturbers.
Even though these new trajectory terms are divided by 100 (the
number of realizations), the remaining effects create small
deviations.

We have demonstrated that the ensemble average technique reveals the
nonuniversal properties of the system, and short-trajectory
corrections can help to better describe the nonuniversal part than
using the radiation impedance alone. Now we demonstrate the benefit
of using the short-trajectory correction in another respect. That
is, when removing the nonuniversal part to reveal the universal
properties, the revealed universal properties have better agreement
with the prediction of RMT if we take account of the
short-trajectory correction. We compare the probability distribution
of the eigenvalues of the normalized impedance $i\xi_{0}$
[normalized by the radiation impedance (see Eq.
(\ref{eq:normalized_impedance}))], $i\xi^{(L_{M})}$ [normalized by
the short-trajectory-corrected impedance (see Eq. (\ref{eq:SOC
normalized impedance}))], and the PDFs that were generated from
numerical RMT \cite{S1, S3, SamThesis}.

\begin{figure}
\includegraphics[scale=0.4]{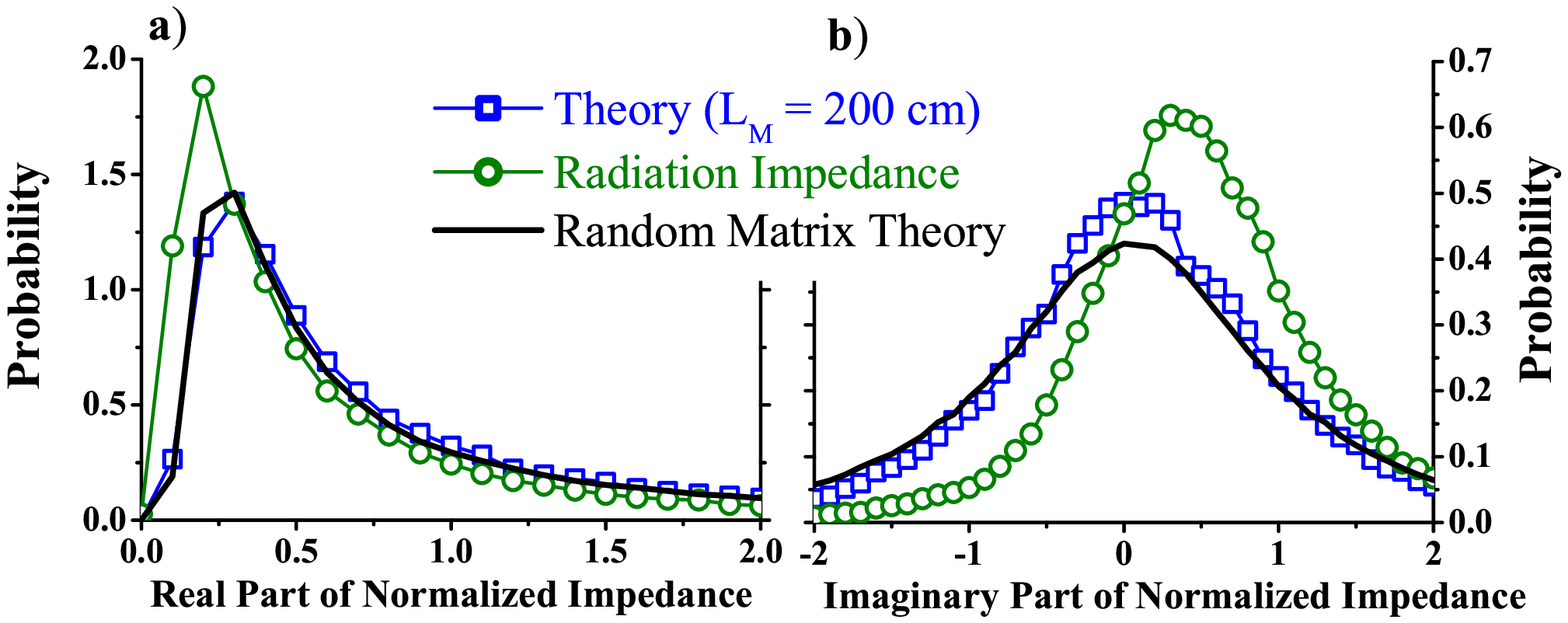}
\includegraphics[scale=0.4]{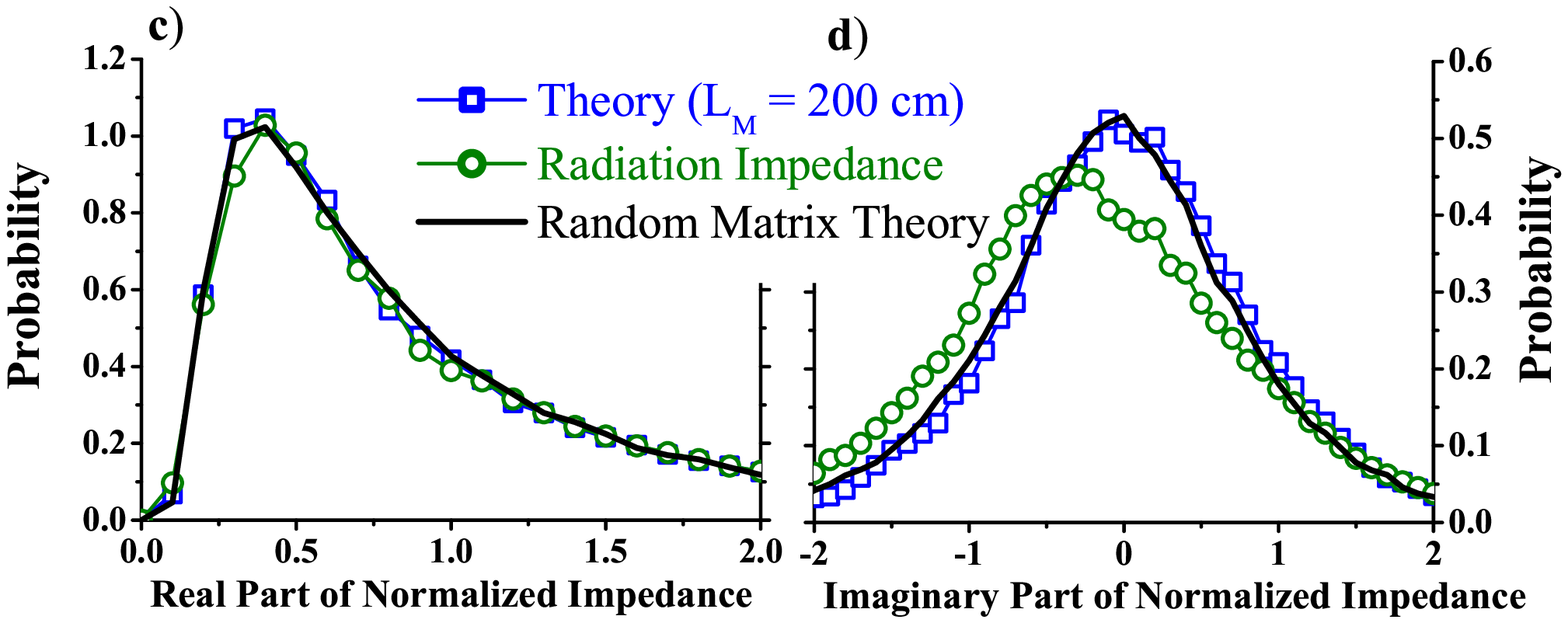}
\caption{(Color online) Plot of probability distributions of (a) the
real part and (b) the imaginary part of the normalized impedance
eigenvalues in the frequency range 6.8 GHz $\sim$ 7.0 GHz, and (c)
the real part and (d) the imaginary part of the normalized impedance
in the frequency range 11.0 GHz $\sim$ 11.2 GHz. The blue (squares)
curve is $i\xi^{(L_{M})}$ that is normalized with the
short-trajectory-corrected impedance, the green (circles) curve is
$i\xi_{0}$ that is normalized with the radiation impedance only, and
the black curve shows the PDF generated from numerical RMT.}
\label{fig:PDF_normalized_impedance}
\end{figure}

Fig. \ref{fig:PDF_normalized_impedance} shows the PDFs of the
normalized impedance $i\xi_{0}$, $i\xi^{(L_{M})}$, and the
corresponding data from numerical RMT. For $i\xi_{0}$ in Eq.
(\ref{eq:normalized_impedance}), we take the impedance $Z$ from the
100 realizations and in a frequency range of 200 MHz. Similarly, for
$i\xi^{(L_{M})}$ in Eq. (\ref{eq:SOC normalized impedance}), we use
the same measured data but consider short-trajectory correction with
$L_{M}=200$ cm in addition to the radiation impedance. Here we show
two examples for frequency ranges 6.8 GHz $\sim$ 7.0 GHz in Figs.
\ref{fig:PDF_normalized_impedance} (a) and (b); and 11.0 GHz $\sim$
11.2 GHz in Figs. \ref{fig:PDF_normalized_impedance} (c) and (d). It
should be noted that in the past such narrow frequency ranges were
not used because short trajectories created strong deviations from
RMT predictions \cite{S1, S2, S3}. Because the normalized impedance
is complex, Figs. \ref{fig:PDF_normalized_impedance} (a) and (c)
show the distribution of the real part of the normalized impedance;
(b) and (d) are the imaginary part. For the numerical RMT data, the
loss parameters ($\alpha = 0.3$ for 6.8 GHz $\sim$ 7.0 GHz and
$\alpha = 0.4$ for 11.0 GHz $\sim$ 11.2 GHz) were determined by the
best matched distribution with a much wider frequency range (2 GHz).
As seen in Fig. \ref{fig:PDF_normalized_impedance} the distribution
of the normalized impedance has much better agreement with the
prediction of RMT when we consider the short-trajectory correction
up to the trajectory length of 200 cm.

The deviations of the distribution between the normalized impedance
and the predictions of RMT in Fig.
\ref{fig:PDF_normalized_impedance} are due to short trajectories
that remain in the ensemble average of the 100 realizations. We have
seen that the PDFs of $i\xi_{0}$ and $i\xi^{(L_{M})}$ approach each
other when we use the data in a wider frequency range. This is
because in a wide enough frequency window the fluctuations in the
impedance due to a trajectory can be compensated (i.e., the required
window is 1.8 (GHz) $\simeq c/17$(cm) for the shortest trajectory
with $L_{b(1,1)}=15$ cm and $L_{port(1,1)}=2$ cm). Therefore, we
take a 2 GHz frequency window to obtain a universal distribution
that is independent of which normalization methods we use (Eq.
(\ref{eq:normalized_impedance}) or Eq. (\ref{eq:SOC normalized
impedance})), and we also determine the loss parameter from this
universal distribution \cite{S1, S3, SamThesis}.

Furthermore, the deviations shown in Fig.
\ref{fig:PDF_normalized_impedance} match the difference between
$\langle Z\rangle$ and $Z_{R}$ shown in Fig.
\ref{fig:ensemble_impedance}. For example, in the frequency range
6.8 GHz $\sim$ 7.0 GHz the ensemble averaged impedance $\langle
Z_{11}\rangle$ is smaller than the radiation impedance $Z_{R,11}$ in
the real part and larger in the imaginary part, and in Figs.
\ref{fig:PDF_normalized_impedance} (a) and (b) we can see the same
bias of the distribution of the normalized impedance $i\xi_{0}$ that
is normalized with the radiation impedance only. Therefore, with
short-trajectory corrections, we can better explain the deviations
between the measured ensemble average and the universal properties
predicted by RMT.

\subsection{Uncovering RMT statistics of the scattering matrix}

\begin{figure}
\includegraphics[scale=0.6]{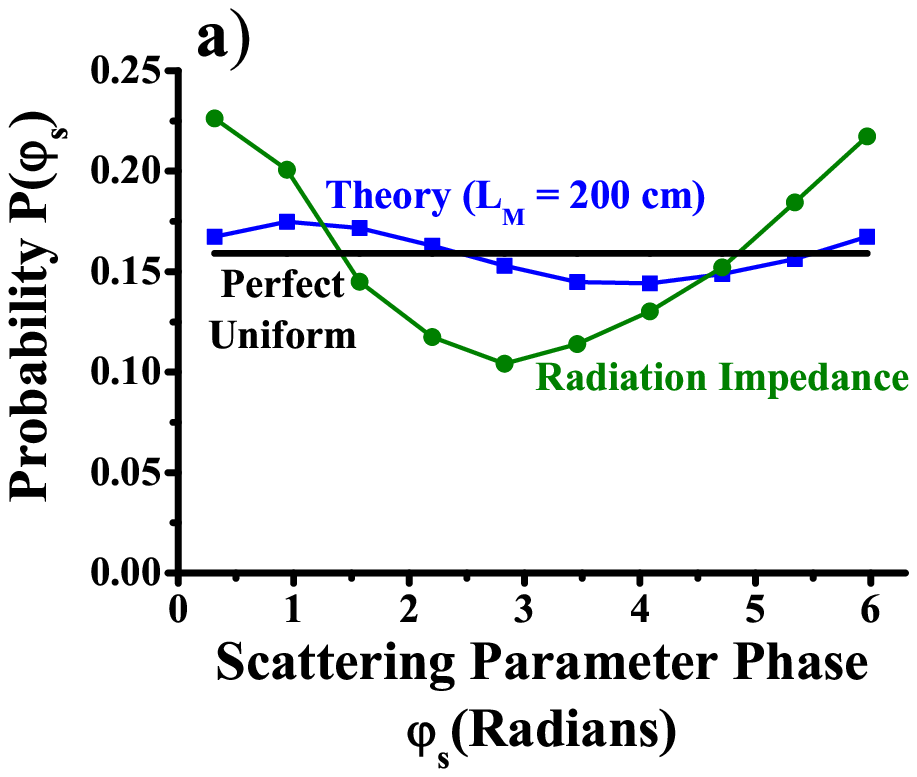}
\includegraphics[scale=0.7]{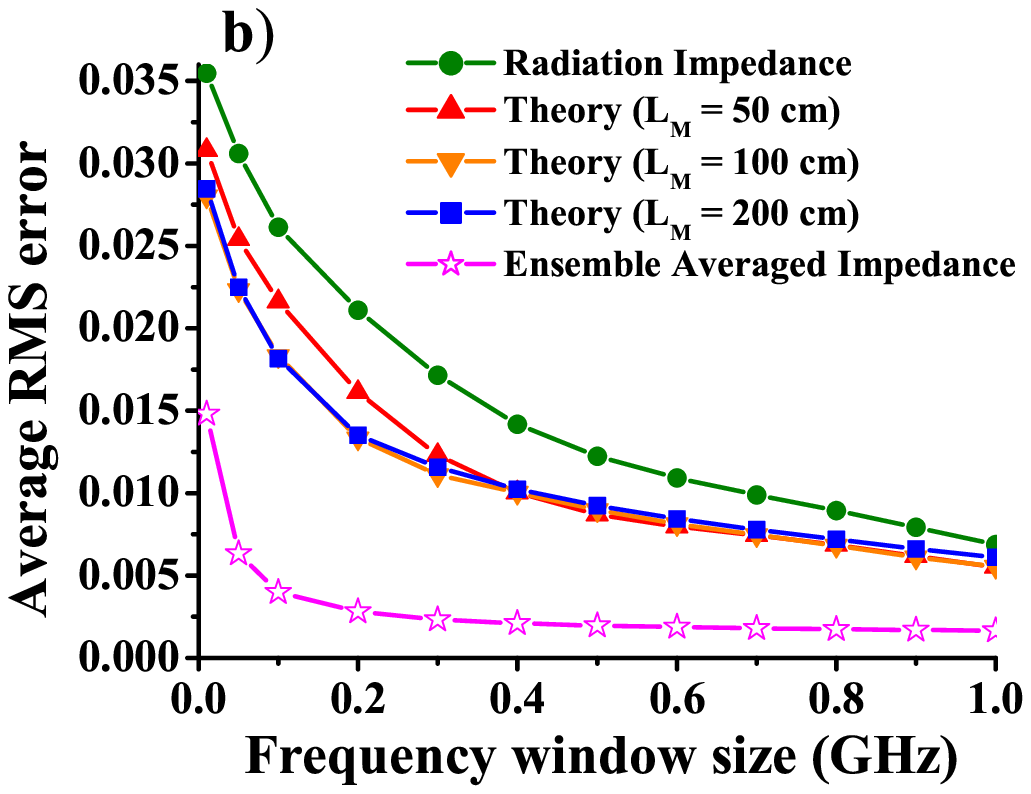}
\caption{(Color online) Plot of (a) an example probability
distribution of the phase $\varphi_{s}$ of eigenvalues of the
normalized scattering matrix $s^{(L_{M})}$ from $0\sim2\pi$, taken
over the 11$\sim$11.5 GHz frequency window inside the $1/4$-bow-tie
cavity of 100 realizations with perturbers. The blue (squares) curve
is from data normalized with $L_{M} = 200$ cm, the green (circles)
curve is from data normalized with the radiation impedance only, and
the black line shows the perfectly uniform distribution for
comparison. Plot of (b) the average RMS error of distributions of
$\varphi_{s}$, where the normalized scattering matrix $s^{(L_{M})}$
is calculated from impedance normalized with the radiation impedance
only (green circles), with the measured ensemble averaged impedance
(pink stars), or with theoretical impedance with $L_{M} = 50$ cm
(red triangles), 100 cm (orange invert triangles), up to 200 cm
(blue squares), versus frequency window size from 10 MHz to 1.0
GHz.} \label{fig:S_statistics}
\end{figure}

We now test the benefits of the short-trajectory correction in
uncovering universal RMT statistics of the scattering matrix. We
find that including short trajectories in the impedance
normalization improves our determination of the RMT statistical
properties of the scattering matrix $S$, given the same amount of
ensemble and frequency averaging. In other words, we show that the
need to resort to wavelength averaging over large numbers of modes
is significantly reduced after including short-trajectory
corrections in the impedance normalization. This section shows the
result of the two-port experiment; the result of the one-port
experiment has been published in our previous work \cite{Jenhao}.
Here we examine the phase $\varphi_{s}$ of the eigenvalues of the
normalized scattering matrix $s^{(L_{M})}$ because the statistics of
$\varphi_{s}$ do not change with the frequency-dependent loss
parameter $\alpha(f)$ \cite{S3}. The normalized scattering matrix
$s^{(L_{M})}$ is defined as
$(i\xi^{(L_{M})}-1)/(i\xi^{(L_{M})}+1)=|s^{(L_{M})}|e^{i\varphi_{s}}$,
where the normalized impedance $i\xi^{(L_{M})}$ is given by Eq.
(\ref{eq:SOC normalized impedance}) with corrections by Eqs.
(\ref{eq:real part SOC}) and (\ref{eq:imaginary part SOC}). Note for
the two-port experiment, we compute the eigenvalues of the matrix
$s^{(L_{M})}$, and $\varphi_{s}$ is the phase of the eigenvalues.
RMT predicts that $\varphi_{s}$ should have a uniform distribution
from 0 to $2\pi$ independent of loss \cite{Poisson_Kernel_Original,
Brouwer_Lorentzian}, as verified in previous experiments \cite{S2,
S3}.

Fig. \ref{fig:S_statistics} illustrates the benefits of using
short-trajectory-corrected data to examine the statistical
properties of the scattering matrix. An ensemble of data is created
by averaging 100 realizations of the cavity with the two perturbers
present, between 6 and 18 GHz, encompassing about 1070 closed-cavity
modes. Theoretical short trajectory corrections are weighted by the
survival probability ($p_{b(n,m)}$) which are calculated from the
known locations of the perturbers.  Fig. \ref{fig:S_statistics}(a)
shows a representative probability distribution ($P(\varphi_{s})$)
of the phase of eigenvalues of the normalized scattering matrix. The
data for the normalized scattering matrix are formed by taking the
measured impedance from each of the 100 realizations and normalizing
with the radiation impedances $Z_{R}$ or the theoretical impedance
with short-trajectory correction $Z_{avg}^{L_{M}}$, and these PDFs
are computed from the data in a 500 MHz frequency range
($11.0\sim11.5$ GHz). In this frequency range, it is clear that the
distribution normalized with short trajectories is significantly
closer to a uniform distribution than the one normalized without
short trajectories.

To quantify this improvement, it is convenient to look at the
average root-mean-square (RMS) error with respect to the uniform
distribution for the two types of normalization (Fig.
\ref{fig:S_statistics}(b)). Here the RMS error is defined as
\begin{equation}\label{eq:RMS}
    RMS\ error\equiv\sqrt{\frac{1}{10}\sum_{i=1}^{10}\left(\frac{n_{i}}{\langle n_{i}\rangle}-1\right)^{2}},
\end{equation}
where $n_{i}$ is the number of elements in the $i^{th}$ bin in the
10 bins histogram of $P(\varphi_{s})$, and $\langle n_{i}\rangle$ is
the mean of $n_{i}$. Therefore, when a distribution is more uniform,
its RMS error is smaller. The average RMS error of each frequency
window is averaged over the spectral range from 6 to 18 GHz, and
Fig. \ref{fig:S_statistics}(b) shows the average RMS error versus
the frequency window sizes from 10 MHz to 1.0 GHz. This figure shows
that the distributions of data are systematically more uniform as
more short trajectories are taken into account in the normalization
for a given window size. The average RMS error of the data
normalized without short trajectories (the radiation impedance only
(green circles)) has the largest error. When we include more short
trajectories from the maximum lengths $L_{M} = 50$ cm (red
triangles) up to $L_{M} = 200$ cm (blue squares), the average RMS
errors decrease. The improvement is obvious after including short
trajectories with $L_{M} = 50$ cm, and it saturates for trajectories
with $L_{M} = 100$ cm. For comparison, we also add a curve (pink
stars) for the data normalized by the measured ensemble averaged
impedance $\langle Z\rangle$ (the red curves in Fig.
\ref{fig:ensemble_impedance}), and it is the most uniform case. It
is observed that the improvement of statistical properties with
short-trajectory corrections is more significant when the frequency
window size is smaller.

\section{Conclusions}
A theory for the nonuniversal effects of coupling and short ray
trajectories on scattering wave-chaotic systems has been tested
experimentally on a two-dimensional, wave-chaotic, electromagnetic
cavity with one or two transmission line channels connecting to the
outside. In particular, the theoretical predictions match the
measured data in the cases of a frequency ensemble in a single
realization and the configuration ensemble in the ensemble of 100
realizations. By removing nonuniversal effects from measured data,
we can reveal underlying universally fluctuating quantities in the
scattering and impedance matrices. These results should be useful in
many fields where similar wave phenomena are of interest, such as
nuclear scattering, atomic physics, quantum transport in condensed
matter systems, electromagnetics, acoustics, geophysics, etc.

\section*{Acknowledgements}
We thank Sameer Hemmady and Florian Schaefer for help with
preliminary experiments and analysis, and R. E. Prange and Michael
Johnson for assistance with theory. The work is funded by ONR grant
N00014-07-1-0734, NSF, AFOSR grant FA95500710049, and ONR grant
N00014-09-1-1190.


\begin{references}
\bibitem{impedance} The impedance matrix gives the linear relation
between the voltages at the ports and the currents. For the problem
we consider, the ports are antennas that are much smaller than a
wavelength and thus, have fixed current distributions. In this case,
if the current at a port/antenna is specified, a wave incident on it
from the interior will not be scattered. Thus, contributions to the
voltage at that port, due to ray trajectories leading to that port,
may be summed. Scattering of waves by the port/antenna occurs when
the port current is allowed to respond to the port voltage, for
example, if the port is connected to a transmission line with
characteristic impedance $Z_{0}$. Note, the effect of multiple
refections from the ports is included in the scattering matrix
$S=(Z-Z_{0})/(Z+Z_{0})$.
\end{references}

\end{document}